\documentclass[aps,pra,reprint,floatfix]{revtex4-1}

\usepackage[T1]{fontenc}

\usepackage{graphicx,amsmath,amssymb,xr}

\newcommand{\rs}{r_\mathrm{s}}

\newcommand{\kf}{k_\mathrm{F}}

\newcommand{\suc}{_\mathrm{c}}
\newcommand{\sux}{_\mathrm{x}}
\newcommand{\suxc}{_\mathrm{xc}}
\newcommand{\nup}{n_\uparrow}
\newcommand{\ndn}{n_\downarrow}

\begin{document}

\title{QMC-consistent static spin and density local field factors for the uniform electron gas}
\author{Aaron D. Kaplan}
\affiliation{Department of Physics, Temple University, Philadelphia, PA 19122}
\email{Currently affiliated with Lawrence Berkeley National Laboratory, Berkeley, CA 94720. ADKaplan@lbl.gov}
\author{Carl A. Kukkonen}
\affiliation{33841 Mercator Isle, Dana Point, CA 92629}
\email{kukkonen@cox.net}
\date{\today}

\begin{abstract}
  Analytic mathematical models for the static spin ($G_-$) and density ($G_+$) local field factors for the uniform electron gas (UEG) as functions of wavevector and density are presented.
  These models closely fit recent quantum Monte Carlo (QMC) data and satisfy exact asymptotic limits.
  This model for $G_-$ is available for the first time, and the present model for $G_+$ is an improvement over previous work.
  The QMC-computed $G_\pm$ are consistent with a rapid crossover between theoretically-derived small-$q$ and large-$q$ expansions of $G_\pm$.
  These expansions are completely determined by $\rs$, the UEG correlation energy per electron, and the UEG on-top pair distribution function.
  We demonstrate their utility by computing uniform electron gas correlation energies over a range of densities.
  These models, which hold over an extremely wide range of densities, are recommended for use in practical time-dependent density functional theory calculations of simple metallic systems.
  A revised model of the spin susceptibility enhancement is developed that fits QMC data, and does not show a ferromagnetic instability at low density.
\end{abstract}

\maketitle

A critical quantity for evaluating the linear response of an interacting uniform electron gas (UEG), or simple metal, are the local field factors (LFFs) $G_\pm(\rs,q,\omega)$.
The UEG (sometimes called jellium) can be characterized by a Wigner-Seitz density parameter $\rs = [3/(4\pi n)]^{1/3}$ and relative spin-polarization $\zeta = (\nup - \ndn)/n$, for total density $n = \nup + \ndn$.
The density (spin-symmetric) LFF $G_+(\rs,q,\omega)$ governs the density-density response $\chi(q,\omega)$ of a many-electron density to a wavevector $q$- and frequency $\omega$-dependent perturbation via \cite{giuliani2005}
\begin{equation}
  \chi^{-1}(q,\omega) = \chi_0^{-1}(q,\omega)
  - \frac{4\pi}{q^2}\left[1 - G_+(\rs,q,\omega) \right].
\end{equation}
$\chi_0(q,\omega)$ is the response function of non-interaction electrons; for the UEG, this is the Lindhard function \cite{lindhard1954}.
Thus $G_+$ is related to the exchange-correlation kernel $f\suxc$ of time-dependent density functional theory \cite{runge1984,gross1985} as $G_+(\rs,q,\omega) = -q^2 f\suxc(\rs,q,\omega)/(4\pi)$.
The spin (antisymmetric) LFF governs the paramagnetic spin-response via \cite{giuliani2005}
\begin{equation}
  \chi_{S_z S_z}^{-1}(q,\omega) = \chi_0^{-1}(q,\omega)
  + \frac{4\pi}{q^2} G_-(\rs,q,\omega).
\end{equation}
There exist many approximate expressions of $G_+$ or $f\suxc$, which range from those which are local in space and time \cite{zangwill1980}, nonlocal in space only (as in this work) \cite{corradini1998}, nonlocal in time only \cite{gross1985,qian2002}, or nonlocal in both space and time \cite{richardson1994,ruzsinszky2020,kaplan2022}.
However, there are no \emph{realistic} expressions of $G_-$ other than that of Richardson and Ashcroft (RA) \cite{richardson1994}, which is based on perturbation theory calculations, and is complicated by typographical errors.
As we make extensive comparisons to the RA LFFs, we correct these typographical errors in Supplemental Material Sec. \ref{sec:RA_corrected}.
The RA LFFs are presumably most realistic at higher densities typical of simple metals, and less realistic at lower densities.

This work provides flexible, analytic expressions for the static LFFs $G_\pm(\rs,q)\equiv \lim_{\omega \to 0} G_\pm(\rs,q,\omega)$ based on known asymptotic limits.
Free parameters are then fitted to recent variational diagrammatic quantum Monte Carlo (QMC) calculations \cite{kukkonen2021}.
This QMC data covers the region below $q = 2.34 \kf$ for $\rs =1-5$ for $G_-$, but is only available for $\rs=1$ \& 2 for $G_+$.
The current model of $G_+(\rs,q)$ also more reliably fits older QMC data \cite{moroni1995} that covers $\rs = 2, 5,$ \& 10, but with no data below $\kf$, than the expression due to Corradini \textit{et al.} \cite{corradini1998}, and provides accurate predictions of the UEG correlation energy.

Both $G_\pm(\rs,q)$ are characterized by a rapid crossover between small- and large-$q$ asymptotics near $q = 2 \kf$, with $\kf = (3\pi^2 n)^{1/3}$ the Fermi wavevector.
This crossover is likely responsible for the ``$2\kf$-hump'' phenomenon \cite{overhauser1970,utsumi1980}: a maximum in $G_+(q)$ may exist for $q \approx 2 \kf$.
The presence of a peak can markedly change the properties of phonon dispersion \cite{wang1984}, superconducting critical temperatures \cite{shirron1986}, etc. when using $G_+(q)$ to approximate the LFF of simple metals in TD-DFT.
Moreover, explicit inclusion of the spin-dependence of the electronic response via $G_-$ is crucial for describing pairing of electrons in superconducting phases \cite{kukkonen1979,buche1990}.
Thus a realistic approximation of $G_-$ at all possible densities and wavevectors is needed to understand the spin-dependence of the electronic response.
Such a model $G_-$ would enable realistic calculations of simple metals using the Kukkonen-Overhauser framework \cite{kukkonen1979} or other theories of linear response.

In this brief paper, we present the formulas for $G_+$ and $G_-$ for all wave vectors given only the density $\rs$.
The details of the curve fitting, asymptotic behavior, and code are given in the Supplemental Material.
The formulas may look complex, but are simple to implement computationally; a documented Python implementation is provided in the public code repository \cite{code_repo}.
More, the models with optimized parameters can be accessed from PyPI by pip installing ``AKCK\_LFF.''

The QMC data for both $G_+$ and $G_-$ closely follow the theoretical asymptotic behavior of varying as $q^2$ at small $q$.
The coefficients of $q^2$ are determined by the compressibility and susceptibility sum rules.
The QMC data rises somewhat faster than $q^2$ to about $2\kf$, and then falls rapidly.
Theory predicts that the large-$q$ behavior of $G_\pm$ is $B_\pm + C q^2$.
Although $B_+$ and $B_-$ differ, they are determined by $\rs$ and the on-top pair correlation function.
$C$ is the same for both $G_\pm$.
The qualitatively similar behaviors of the LFFs permit us to use the same analytically simple expressions, defined below in Eqs. (\ref{eq:g_app}) and (\ref{eq:alpha_fit}), to model $G_+$ and $G_-$.

The fitting process, partially described below, simply allows the small-$q$ behavior to rise above $q^2$, combined with an adjustable exponential cutoff near $2\kf$.
This cutoff modulates the transition to the large $q$ asymptotics.
The recent QMC data stops at $2.34 \kf$, but is consistent with the large-$q$ asymptotic behavior, assuming a simple transition.
The following equations completely specify the local field factors.

Let $x \equiv q/\kf$, then we model \textit{both} $G_\pm$ as
\begin{align}
  & G_j(\rs,q) =
    x^2\left[A_j(\rs) + \alpha_j(\rs) x^4 \right]
    H\left(x^4/16; a_{3j}, a_{4j} \right)
    \nonumber \\
  & + \left[C(\rs)x^2 + B_j(\rs) \right]
    \left[ 1 - H\left(x^4/16; a_{3j}, a_{4j} \right) \right],
    \label{eq:g_app} \\
  & \alpha_j(\rs) = a_{0j} + a_{1j} \exp(-a_{2j} \rs),
    \label{eq:alpha_fit}
\end{align}
where $j = +, \, -$.
The smoothed step function
\begin{equation}
  H(y; \beta,\gamma) = \frac{\left(e^{\beta \gamma} -1 \right)e^{-\beta y}}{1 + \left(e^{\beta \gamma} - 2 \right)e^{-\beta y} }
\end{equation}
is constructed to satisfy three limits: $H(0;\beta,\gamma) = 1$; $H(\gamma;\beta,\gamma)=1/2$; and $H(\infty;\beta,\gamma)=0$.
While $H$ has no physical basis, it represents a simple and reasonable transition from the low-$q$ behavior of the QMC data to the large-$q$ asymptotics.
The $a_{ij}$ parameters are fitted to QMC data.

Equation (\ref{eq:g_app}) satisfies the exact small-$q$ expansions (SQEs) of $G_\pm$, which are identical in structure.
For $G_+$, this is the compressibility sum rule:
\begin{align}
  \lim_{q \to 0} G_+(\rs,q) &= A_+(\rs) x^2 + \mathcal{O}(x^4), \label{eq:gp_small_q} \\
  A_+(\rs) &= -\frac{\kf^2}{4\pi} \frac{\partial^2 e\suxc^\text{LDA}}{\partial n^2}(\rs),
\end{align}
with $e\suxc^\text{LDA}$ the local-density approximation \cite{dirac1930,kohn1965,perdew1992} for the UEG exchange-correlation energy density.
Unless specified, we use Hartree atomic units, $\hbar = m_e = e^2 =1$; 1 Hartree energy unit is 2 Rydberg, 27.211386 eV; 1 bohr length unit is 0.529177 \AA{} \cite{unit_data}.
The SQE of $G_-$ is the susceptibility sum rule \cite{giuliani2005}:
\begin{align}
  \lim_{q \to 0} G_-(\rs,q) &= A_-(\rs) x^2 + \mathcal{O}(x^4),
  \label{eq:gm_small_q} \\
  A_-(\rs) &= - \frac{3 \pi}{4 \kf}\frac{\partial^2 \varepsilon^\text{LSDA}\suxc}{\partial \zeta^2}(\rs,0).
  \label{eq:a_minus}
\end{align}
For simple polynomial approximations of $A_\pm(\rs)$ valid for $1 \leq \rs \leq 5$, see Eqs. (6) and (7) of Ref. \cite{kukkonen2021}.
$\varepsilon^\text{LSDA}\suxc$ is the local spin-density approximation for the UEG exchange-correlation energy per electron, for which we use the Perdew-Wang approximation \cite{perdew1992}.
The quantity
\begin{equation}
  \alpha\suxc(\rs) \equiv \frac{\partial^2 \varepsilon^\text{LSDA}\suxc}{\partial \zeta^2}(\rs,0)
\end{equation}
is often called the spin-stiffness \cite{vosko1980}.
The exchange contribution to the spin-stiffness can be shown to be $\alpha\sux(\rs) = -\kf/(3\pi)$ \cite{dirac1930,kohn1965,oliver1979}.

Equation (\ref{eq:g_app}) also satisfies the large-$q$ expansions (LQEs) of $G_\pm$, again identical in structure.
For $G_+$, \cite{corradini1998}
\begin{align}
  \lim_{q \to \infty} G_+(\rs,q) &= C(\rs) x^2 + B_+(\rs) + \mathcal{O}(x^{-2}),
  \label{eq:gp_large_q} \\
  C(\rs) &= -\frac{\pi}{2\kf} \frac{\partial}{\partial \rs}
  \left[ \rs \varepsilon^\text{LDA}\suc(\rs)\right].
\end{align}
The function $B_+(\rs)$ is parameterized as \cite{moroni1995}
\begin{equation}
  B_+(\rs) = \frac{1 + (2.15) \rs^{1/2} + (0.435) \rs^{3/2}}{3 + (1.57) \rs^{1/2} + (0.409) \rs^{3/2}}.
\end{equation}
The LQEs of $G_-$ and $G_+$ are connected as \cite{richardson1994,niklasson1974,zhu1984,giuliani2005}
\begin{align}
  \lim_{q\to \infty} G_-(\rs,q) &= C(\rs) x^2 + B_-(\rs) + \mathcal{O}(x^{-2}),
  \label{eq:gm_large_q} \\
  B_-(\rs) &= B_+(\rs) + 2g(\rs) - 1, \label{eq:bminus}
\end{align}
i.e., they differ only by the on-top pair distribution function $g(\rs)$, which we approximate as \cite{perdew1992b}
\begin{equation}
  g(\rs) = \frac{1}{2} \frac{1 + 2 (0.193) \rs}{\{1 + (0.525) \rs[1 + (0.193) \rs]\}^2}.
\end{equation}

\begin{table*}
  \centering
  \begin{tabular}{rrrr} \hline
    $j=$ & $+$ ($G_+$) & $-$ ($G_-$) & $-$ ($G_-$), new $\alpha\suc$ \\ \hline
    $a_{0j}$ & $-0.00451760 \pm 0.002$ & $-0.00105483 \pm 0.0008$ & $-0.000519869 \pm 0.0008$ \\
    $a_{1j}$ & $0.0155766 \pm 0.002$ & $0.0157086 \pm 0.0006$ & $0.0153111 \pm 0.0005$ \\
    $a_{2j}$ & $0.422624 \pm 0.2$ & $0.345319 \pm 0.05$ & $0.356524 \pm 0.05$ \\
    $a_{3j}$ & $3.516054 \pm 0.5$ & $2.850094 \pm 0.1$ & $2.824663 \pm 0.1$ \\
    $a_{4j}$ & $1.015830 \pm 0.04$ & $0.935840 \pm 0.02$ & $0.927550 \pm 0.02$ \\ \hline
  \end{tabular}
  \caption{Fit parameters $a_{ij}$ for the model LFFs of Eq. (\ref{eq:g_app}) and the estimated uncertainties in the parameters.
  $i = 1, 2, 3, 4$, and $j = +$ for the $G_+$ parameters, and $j=-$ for the $G_-$ parameters.
  The rightmost column uses a revised parameterization for the correlation spin stiffness, described below.
  Only $a_{0-}$ is sensitive to the choice of $\alpha\suc$, although that may be due to its relatively larger  uncertainty.
  }
  \label{tab:fpars}
\end{table*}

To fit Eq. (\ref{eq:g_app}) for $G_\pm$, we minimize the deviation from the QMC-computed values of $G_\pm$, weighted by their corresponding uncertainties.
The fitting method is described fully in Supplemental Material Sec. \ref{sec:fit_method}.
Table \ref{tab:fpars} presents fitted parameters $a_{ij}$ and their uncertainties estimated using a bootstrap method.
This method is described in the Supplemental Material Sec. S1.
We recommend using the full precision of the parameters rather than truncated values based on uncertainty estimates.

\begin{figure}
  \centering
  \includegraphics[width=\columnwidth]{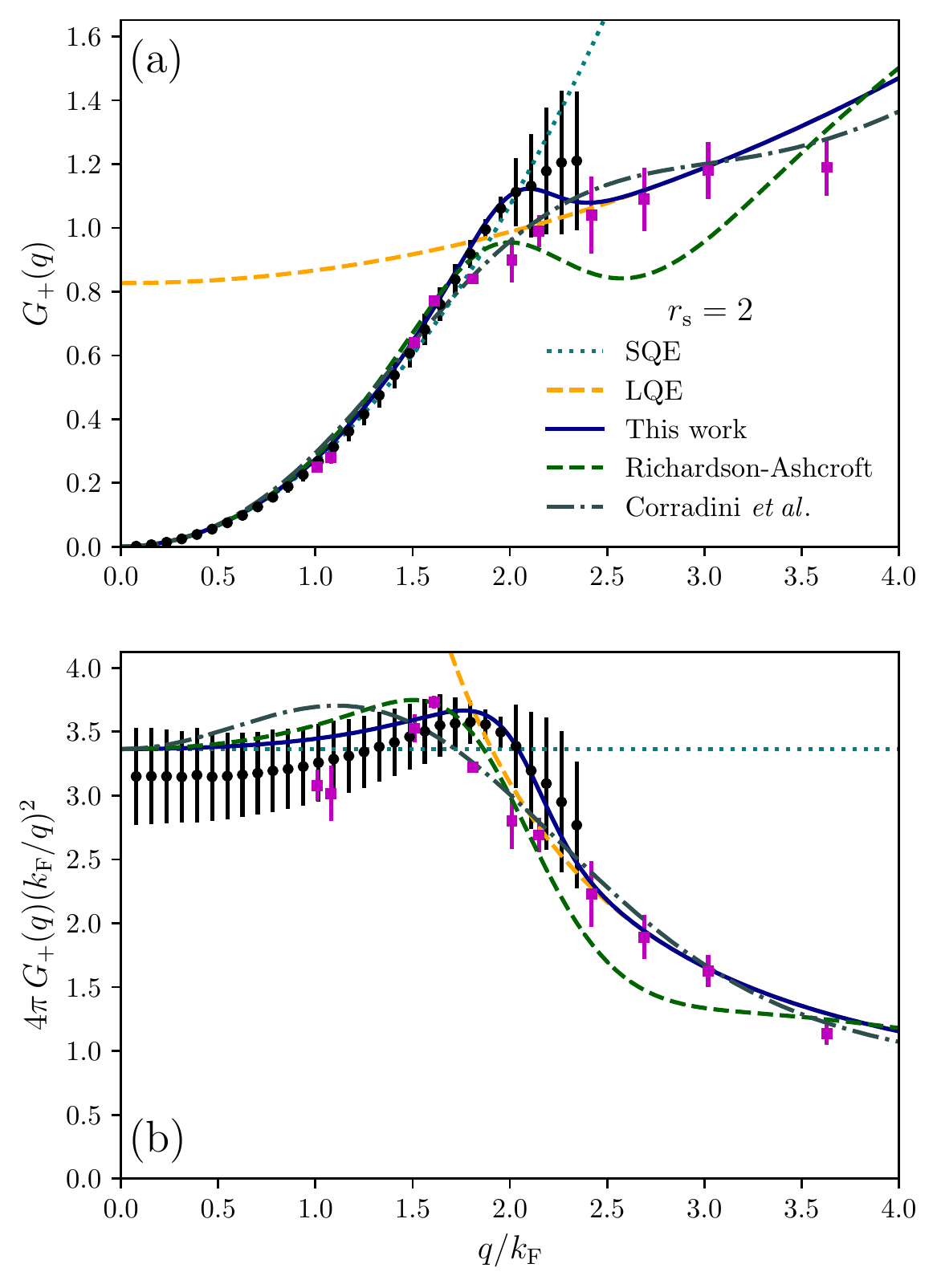}
  \caption{
  Comparison of the model $G_+$ of Eq. (\ref{eq:g_app}) (blue, solid line) and Table \ref{tab:fpars} with the QMC data of Ref. \cite{kukkonen2021} (black circles with vertical uncertainties) and \cite{moroni1995} (magenta squares with vertical uncertainties) for $\rs = 2$.
  Panel (a) presents $G_+$ and (b) $4\pi G_+ (\kf/q)^2 = \kf^2 f\suxc(q)$.
  The latter quantity, essentially the exchange-correlation kernel, is a sensitive test of the fit quality.
  Also shown are the LFFs of Corradini \textit{et al.} \cite{corradini1998} (gray, dash-dotted), which is fitted to the data of Ref. \cite{moroni1995}, and of RA \cite{richardson1994} (green, dashed).
  The small-$q$ expansion (SQE) of Eq. (\ref{eq:gp_small_q}) (teal, dotted) and large-$q$ expansion (LQE) of Eq. (\ref{eq:gp_large_q}) (orange, dashed) are also shown.
  }
  \label{fig:gplus_rs_2}
\end{figure}

Figure \ref{fig:gplus_rs_2} compares our fitted $G_+$ to the data of Ref. \cite{kukkonen2021} and to the older QMC data of Moroni \textit{et al.} \cite{moroni1995} for $\rs = 2$.
The quality of fit is excellent, lying within the uncertainty of the QMC data at all computed points.
The LFF of Corradini \textit{et al.} \cite{corradini1998}, although fitted to the Moroni \textit{et al.} data, fits it poorly.
The LFF developed here, fitted to the Moroni \textit{et al.} data at $\rs = 5$ and 10 only, fits it rather well.

The Supplemental Material presents further plots of $G_+$ that demonstrate the quality of fit to the data of Refs. \cite{kukkonen2021,moroni1995} in Figs. \ref{fig:gp_rs_1}--\ref{fig:gp_rs_10}.
Supplemental Figs. \ref{fig:gp_rs_0p1}--\ref{fig:gp_rs_100} show that our model realistically extrapolates to values of $\rs$ for which there are no QMC data.
For surface plots of $G_+$ at metallic densities, see Figs. \ref{fig:surf_gp_corr} and \ref{fig:surf_gp_ras}.
At a very high density, $\rs = 0.1$ in Fig. \ref{fig:gp_rs_0p1}, our model and the RA $G_+(\rs,q)$ exhibit very similar behaviors: a simple interpolation between small- and large-$q$ asymptotics with a hump near $2\kf$.
At a very low density, $\rs = 100$ in Fig. \ref{fig:gp_rs_100}, our model tends to a smooth, hump-free interpolation between the two regimes, but the RA $G_+$ exhibits likely unphysical oscillations.
This latter behavior of RA is consistent with its derivation from perturbation theory.

Moreover, from Figs. \ref{fig:gplus_rs_2} and \ref{fig:gp_rs_1}--\ref{fig:gp_rs_10}, one can see that the QMC data validates the theoretically-derived asymptotic expansions in the small-$q$ limit, and is also consistent with the large-$q$ limit.
This is direct validation of the compressibility sum rule.
All parameters in $G_+(\rs,q)$ are completely determined by $\rs$ and the UEG correlation energy per electron.

\begin{figure}
  \centering
  \includegraphics[width=\columnwidth]{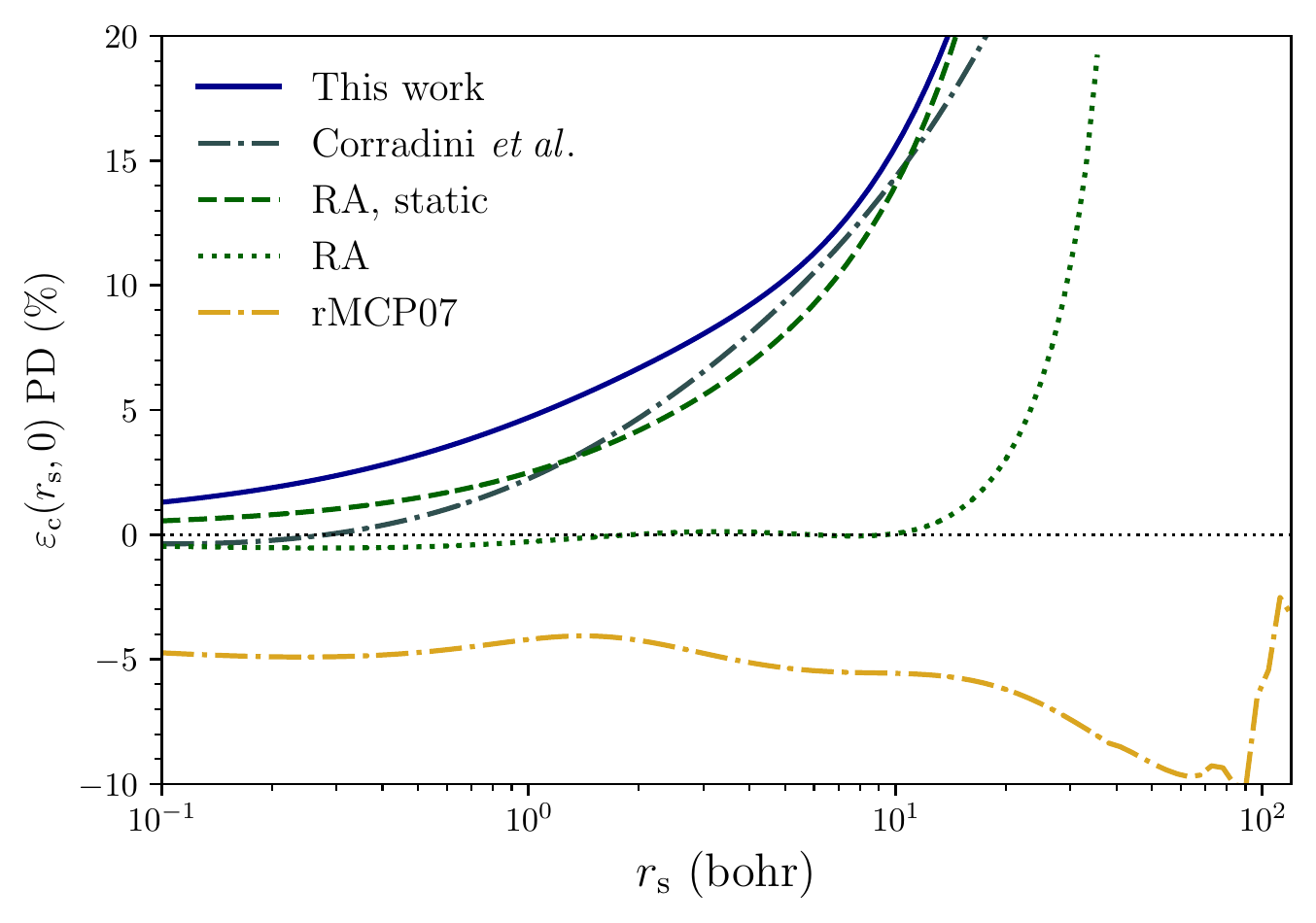}
  \caption{Percent deviation (PD) from the Perdew-Wang approximation \cite{perdew1992} of the UEG correlation energy, using a few common approximations for $G_+$.
  We define the PD as $(100\%)\left[1 - \varepsilon\suc^\text{approx}/\varepsilon\suc^\text{PW92} \right]$.
  The solid blue curve is computed using Eq. (\ref{eq:g_app}) and Table \ref{tab:fpars}.
  The dashed green curve is the static limit of the RA LFF \cite{richardson1994}, and the dotted green curve is its frequency-dependent form.
  The dash-dotted gray curve is due to Ref. \cite{corradini1998}, and the dash-dotted yellow curve to Ref. \cite{kaplan2022}.
  The numeric integration for both variants of the RA LFF appears to become unstable for $\rs \gtrsim 45$.
  }
  \label{fig:eps_c_err}
\end{figure}

Figure \ref{fig:eps_c_err} plots the errors in the UEG correlation energies computed using this model and a few common approximations for $G_+$.
The model of this work systematically overestimates the correlation energies, but makes errors comparable to any of the LFFs presented there.
More accurate correlation energies require a frequency-dependent $G_+(\rs,q,\omega)$, such as those of Refs. \citenum{richardson1994,ruzsinszky2020,kaplan2022}.
The method of computation is described in Supplemental Material Sec. \ref{sec:eps_c_method}, and a validation of our method using the random phase approximation (RPA, $G_+^\text{RPA}=0$) is given in Supplemental Table \ref{tab:RPA_sanity}.

\begin{figure}
  \centering
  \includegraphics[width=\columnwidth]{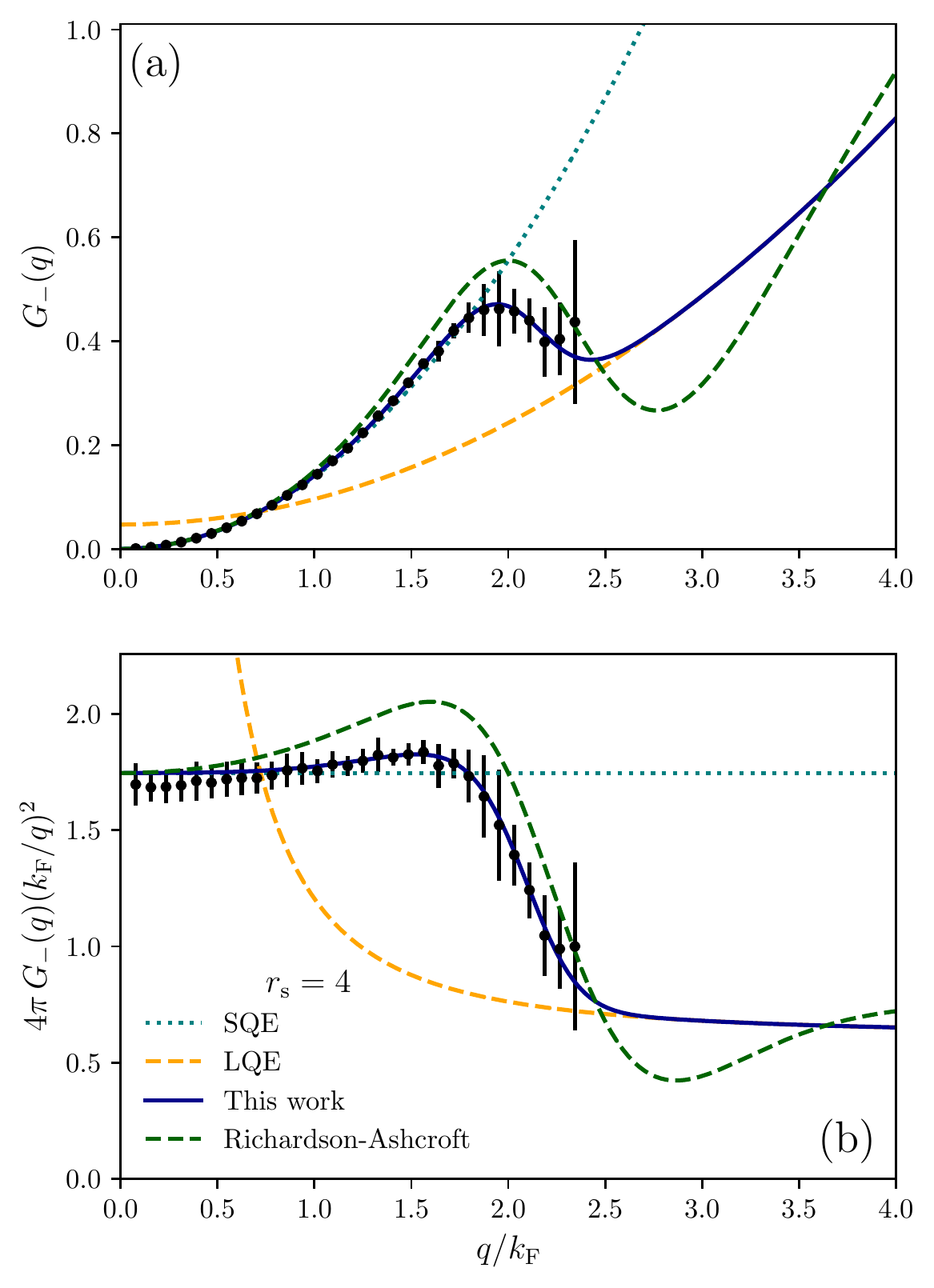}
  \caption{Comparison of the model $G_-$ of Eq. (\ref{eq:g_app}) (blue, solid curve) and Table \ref{tab:fpars} with the QMC data of Ref. \cite{kukkonen2021} (black circles with vertical uncertainties) for $\rs = 4$.
  Panel (a) presents $G_-$ and (b) $4\pi G_-(\kf/q)^2$.
  The static RA \cite{richardson1994} LFF is also shown (green, dashed).
  The small-$q$ expansion (SQE) of Eq. (\ref{eq:gm_small_q}) (teal, dotted) and the large-$q$ expansion (LQE) of Eq. (\ref{eq:gm_large_q}) (orange, dashed) are also shown.
  }
  \label{fig:gminus_rs_4}
\end{figure}

Figure \ref{fig:gminus_rs_4} compares our fitted $G_-$ to the Kukkonen-Chen QMC data \cite{kukkonen2021} for $\rs = 4$.
The quality of fit is again excellent, lying within the QMC uncertainties at all points.
The transition between small- and large-$q$ asymptotics is apparent from Fig. \ref{fig:gminus_rs_4}(b).
Equation (\ref{eq:g_app}) avoids the unusual oscillations present in the RA LFF, which is a rational polynomial in $q^2$.

Supplemental Figs. \ref{fig:gm_rs_1}--\ref{fig:gm_rs_5} demonstrate the high quality of fit to $G_-(\rs,q)$ at other values of $\rs \in \{ 1,2, 3,5\}$.
Extrapolations to the same high, $\rs = 0.1$, and low, $\rs = 100$, densities are made in Figs. \ref{fig:gm_rs_0p1} and \ref{fig:gm_rs_100}, respectively.
The same conclusions regarding $G_+$ hold for $G_-$: our model and RA's are consistent at high densities, but RA's model becomes unphysically oscillatory at low densities.
For surface plots of $G_-$ at metallic densities, see Fig. \ref{fig:surf_gm_ras}.

These figures also show that the QMC data validates the asymptotic expansions of $G_-$, and thus the spin-susceptibility sum rule.
Note that $G_-(\rs,q)$ depends on the parameters of $G_+(\rs,q)$ and the UEG on-top pair distribution function via Eq. (\ref{eq:bminus}).

\begin{figure}
  \centering
  \includegraphics[width=\columnwidth]{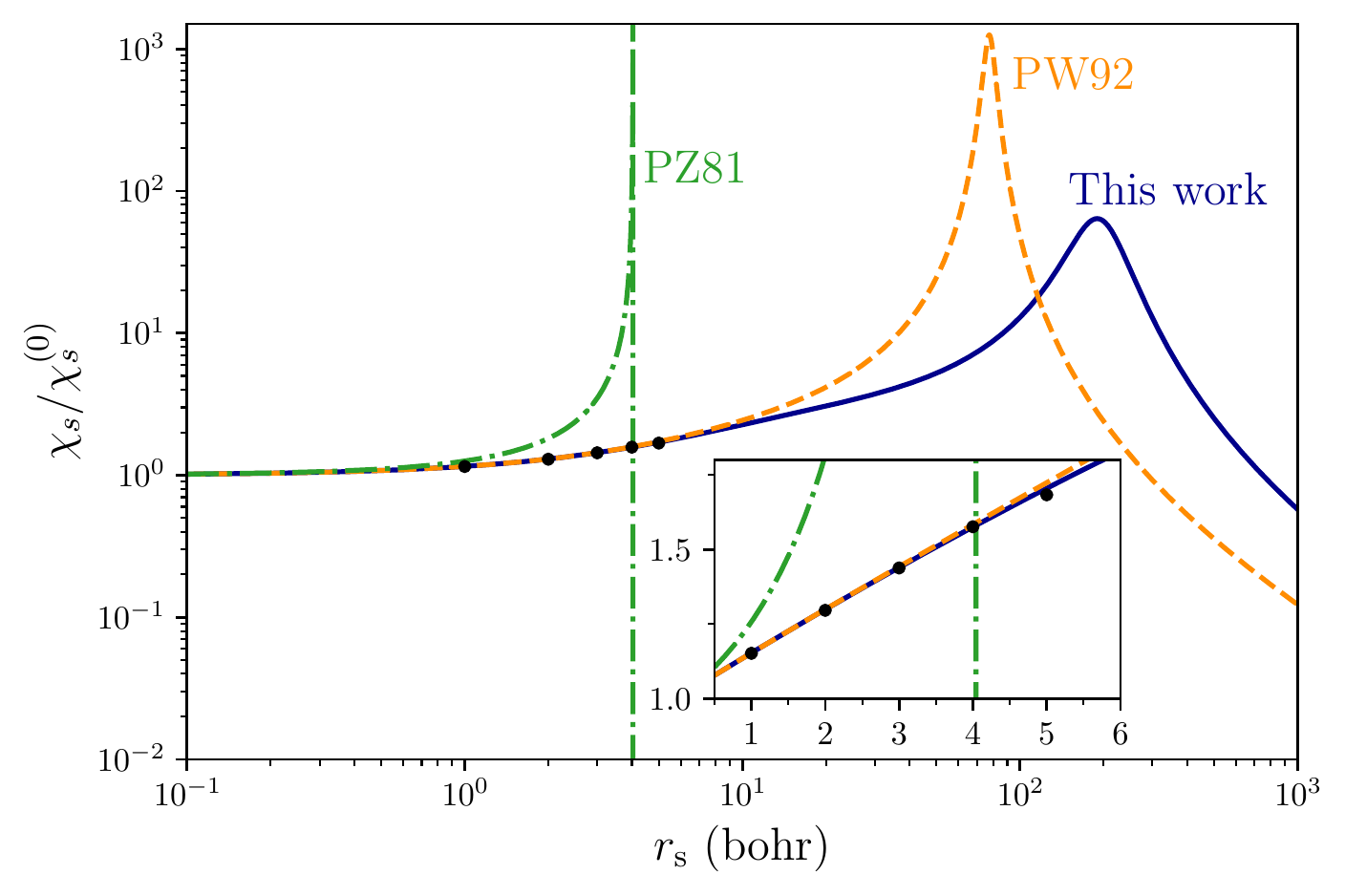}
  \caption{Susceptibility enhancement $\chi_s/\chi_s^{(0)}$ computed with QMC \cite{chen2019,kukkonen2021} (black dots with almost imperceptible error bars), using Eq. (\ref{eq:chi_enh}) with the Perdew-Wang (PW92) \cite{perdew1992} approximation for $\alpha\suc$ (orange, dashed), the re-parameterized form motivated here (blue, solid), or the older expression due to Perdew and Zunger (PZ81) \cite{perdew1981} (green, dash-dotted).
  Although PZ81 includes no explicit information on $\alpha\suc$, it is often used in solid-state and time-dependent density functional calculations.
  The inset shows the range $0.5 \leq \rs \leq 6$.
  }
  \label{fig:chi_enh}
\end{figure}

Last, we discuss the accuracy of the PW92 parameterization of the correlation spin stiffness $\alpha\suc(\rs)$.
It can be observed from either Fig. \ref{fig:chi_enh} or Table \ref{tab:chi_enh} that the enhancement of the interacting spin-susceptibility $\chi_s$, over the non-interacting spin-susceptibility $\chi_s^{(0)}$ (both per unit volume),
\begin{equation}
  \frac{\chi_s}{\chi_s^{(0)}} = \left\{ 1
    - \left( \frac{4}{9\pi} \right)^{1/3} \frac{\rs}{\pi}
    + 3  \left( \frac{4}{9\pi} \right)^{2/3} \rs^2 \alpha\suc(\rs)
  \right\}^{-1}
  \label{eq:chi_enh},
\end{equation}
predicted by PW92 is not consistent with QMC calculations for $\rs > 10$ bohr \cite{chen2019,kukkonen2021}.
For all applications besides low-density jellium, extensive tests have shown PW92 to be robust.
In units of the electron spin moment, $\chi_s^{(0)} = 3n/\kf^2$.
Recent QMC calculations of $\chi_s/\chi_s^{(0)}$ and of the UEG correlation energy at low densities \cite{azadi2022} make it possible to accurately fit $\alpha\suc$ directly.
The Perdew-Wang model of $\alpha\suc(\rs)$ is
\begin{align}
  \alpha\suc(\rs) =& 2 A(1 + \alpha_1 \rs)
    \label{eq:alpha_c_pw92} \\
  & \times \ln \left[1 + \frac{1}{2A(\beta_1 \rs^{1/2}
  + \beta_2 \rs + \beta_3 \rs^{3/2} + \beta_4 \rs^2)} \right],
  \nonumber
\end{align}
where $A$, $\beta_1$, and $\beta_2$ are constrained to ensure the analytic high-density expansion \cite{vosko1980}
\begin{equation}
  \lim_{\rs \to 0} \alpha\suc(\rs) \approx -\frac{1}{6\pi^2} \ln \rs
    + 0.035474401.
\end{equation}
We have recomputed the constant term.
To refit $\alpha\suc$, we minimized the deviation from the tabulated values of the susceptibility enhancement \cite{chen2019,kukkonen2021}, and from approximate values of the spin stiffness at low densities \cite{azadi2022}.
See Supplemental Material Sec. \ref{sec:fit_method} for a description of this method.
Table \ref{tab:alpha_c_rev} presents fitted parameters and expansion coefficients.
Our parameterization is recommended only for applications where higher precision of $\alpha\suc(\rs > 10)$ is needed: our model and PW92 appear to differ at most by about 3.3\% at $\rs = 18.3$ bohr.
We still use the PW92 parameterization of $\alpha\suc$ in our model $G_-$ via Eq. (\ref{eq:a_minus}).
Table \ref{tab:fpars} also provides model parameters for $G_-$ using the current parameterization of $\alpha\suc$.
Consistent with the improvements in $\alpha\suc$, the quality of fit is numerically improved, although the two variants of $G_-$ are visually indistinct.

Consistent with recent QMC-driven analyses of the low-density phases of the UEG \cite{holzmann2020,azadi2022}, our parameterization of $\alpha\suc$ yields no divergence in the susceptibility enhancement.
The present and PW92 parameterizations of $\alpha\suc$ both predict near-divergences in $\chi_s/\chi_s^{(0)}$.
Such a divergence would indicate a ferromagnetic instability in the low-density UEG, whereby a transition from the paramagnetic to ferromagnetic fluid phases is possible.
Both Refs. \cite{holzmann2020} and \cite{azadi2022} find that a transition to a Wigner crystal phase occurs before a transition to the ferromagnetic fluid phase.

In summary, this work presents straightforward analytic models of the static density (spin-symmetric) and spin (antisymmetric) local field factors of the uniform electron gas (UEG), which are fitted to recent QMC data \cite{kukkonen2021}.
These models hold at an extremely wide range of densities, and the model of $G_+$ predicts UEG correlation energies with accuracy sufficient to recommend use in practical calculations of simple metallic systems.
We have also re-parameterized the correlation spin-stiffness of the UEG using QMC data \cite{chen2019,kukkonen2021,azadi2022}, which shows no transition to a ferromagnetic fluid phase.

\begin{table}
  \centering
  \begin{tabular}{rr|rr} \hline
    \multicolumn{2}{c}{$\alpha\suc$ parameter} & \multicolumn{2}{c}{Expansion coefficient}\\ \hline
    $A$ & 0.016886864 & $c_0$ & -0.016886864 \\
    $\alpha_1$ & 0.086888870 & $c_1$ & 0.035474401 \\
    $\beta_1$ & 10.357564711 & $c_2$ & 0.001467281 \\
    $\beta_2$ & 3.623216709 & $c_3$ & 0.005782963 \\
    $\beta_3$ & 0.439233491 & $d_0$ & 0.210976870 \\
    $\beta_4$ & 0.411840739 & $d_1$ & 0.225009568 \\
    \hline
  \end{tabular}
  \caption{Left two columns: parameters appearing in Eq. (\ref{eq:alpha_c_pw92}) for the correlation spin stiffness, $\alpha\suc(\rs)$.
  Right two columns: expansion coefficients derived using these parameters, such that $\lim_{\rs \to 0} \alpha\suc(\rs) = c_0 \ln \rs - c_1 + c_2 \rs \ln \rs - c_3 \rs + ...$ and $\lim_{\rs \to \infty} \alpha\suc(\rs) = -d_0/\rs + d_1/\rs^{3/2} + ...$.
  }
  \label{tab:alpha_c_rev}
\end{table}
\clearpage
\begin{acknowledgments}
  A.D.K. thanks Temple University for a presidential fellowship.
  We acknowledge helpful discussions with John P. Perdew.
\end{acknowledgments}


%

\clearpage

\renewcommand{\thepage}{S\arabic{page}}
\renewcommand{\thesection}{S\arabic{section}}
\renewcommand{\theequation}{S\arabic{equation}}
\renewcommand{\thetable}{S\arabic{table}}
\renewcommand{\thefigure}{S\arabic{figure}}

\onecolumngrid

\section*{Supplemental Material: \\
QMC-consistent static spin and density local field factors for the uniform electron gas}
\twocolumngrid

\tableofcontents

\onecolumngrid

\section{Fitting procedure \label{sec:fit_method}}

To fit the local field factors (LFFs) $G_\pm$, we performed a least squares fit using the SciPy package \cite{virtanen2020}.
The sum of squared residuals
\begin{equation}
  \chi_\pm^2 = \sum_{i,j} \left| \frac{G_\pm(\rs^{(i)},q_j) - G_\pm^\text{QMC}(\rs^{(i)},q_j)}{\delta G_\pm^\text{QMC}(\rs^{(i)},q_j)} \right|^2
\end{equation}
was minimized.
$G_\pm^\text{QMC}$ is the LFF computed from QMC, and $\delta G_\pm^\text{QMC}$ is its uncertainty.
For $G_+$, we fit to the Kukkonen-Chen \cite{kukkonen2021} data for $G_+(0 \leq q_j/\kf < 2.5)$ at $\rs^{(i)} \in \{ 1, 2\} $; and to the Moroni \textit{et al.} \cite{moroni1995} data for $G_+(1 < q_j/\kf < 4.25)$ at $\rs^{(i)} \in \{ 5, 10\}$.
For $G_-$, we fit only to the Kukkonen-Chen \cite{kukkonen2021} data for $G_-(0 \leq q_j/\kf < 2.5)$ at $\rs^{(i)} \in \{ 1, 2, 3, 4, 5 \}$.

To estimate uncertainties in the parameters, we use a ``bootstrap'' method described by Ref. \cite{press1992}.
Suppose we fit to $N$ QMC data points.
From these $N$ data points, we construct $M$ artificial data sets whose contents are randomly selected from the true data set, with replacement.
We then repeat the least squares fit, using the optimal parameters for the true data set as initial guesses; this appeared to be necessary to stabilize the uncertainty estimators.
Call the true, optimized parameters $a_i$.
The parameters from optimization of the $k^\mathrm{th}$ data set will be called $a_i^{(k)}$.
We then compute the mean and variance in the parameters over $M$ synthetic data sets,
\begin{align}
  \overline{a}_i &= \frac{1}{M} \sum_{k=1}^M a_i^{(k)} \\
  \mathrm{var}(a_i) &= \frac{1}{M} \sum_{k=1}^M \left[a_i^{(k)}\right]^2.
\end{align}
The uncertainty in the $i^\mathrm{th}$ parameter is then estimated as
\begin{equation}
  \delta a_i = \left[\mathrm{var}(a_i) - \overline{a}_i^2 \right]^{1/2}.
\end{equation}
In practice, we used $M=1000$ synthetic data sets, and manually inspected their values as a function of increasing $M$ for stability of the uncertainty estimators.

\begin{table}[h]
  \begin{tabular}{rrrrrr} \hline
    & QMC \cite{chen2019,kukkonen2021} & \multicolumn{2}{c}{PW92} & \multicolumn{2}{c}{This work} \\
    $r_\mathrm{s}$ & & $\chi_s/\chi_P$ & PD (\%) & $\chi_s/\chi_P$ & PD (\%) \\ \hline
    1 & 1.152(2) & 1.153425 & 0.12 & 1.153466 & 0.13 \\
    2 & 1.296(6) & 1.299474 & 0.27 & 1.299030 & 0.23 \\
    3 & 1.438(9) & 1.442503 & 0.31 & 1.439717 & 0.12 \\
    4 & 1.576(9) & 1.583653 & 0.48 & 1.575237 & -0.05 \\
    5 & 1.683(15) & 1.723687 & 2.39 & 1.705048 & 1.30 \\
  \hline
  \end{tabular}
  \caption{Values of the spin-susceptibility enhancement calculated in Refs. \cite{chen2019,kukkonen2021}, here by Eq. (\ref{eq:chi_enh}) using the Perdew-Wang (PW92) parameterization \cite{perdew1992} of the UEG correlation energy density, and in this work using a revised parameterization of the Perdew-Wang form.
  The percent difference (PD) in quantities $x$ and $y$ is defined here as $(200\%)(x-y)/(x+y)$, i.e., the difference of $x$ and $y$ weighted by their average.}
  \label{tab:chi_enh}
\end{table}

To fit the correlation spin-stiffness $\alpha\suc$, we performed a least-squares \cite{virtanen2020} minimization of the objective function
\begin{align}
  \sigma =& \sum_i \left|
    \frac{\widetilde{\chi}_\text{approx}(\rs^{(i)}) - \widetilde{\chi}_\text{QMC}(\rs^{(i)})}{\delta \widetilde{\chi}_\text{QMC}(\rs^{(i)})}
    \right|^2
    + \sum_i \left|
    \frac{\alpha\suc(\rs^{(i)}) - \alpha\suc^\text{QMC}(\rs^{(i)})}{\delta \alpha\suc^\text{QMC}(\rs^{(i)})}
  \right|^2.
\end{align}
$\widetilde{\chi}_\text{QMC}=\chi_s^\text{QMC}/\chi_s^{(0)}$, and $\delta \widetilde{\chi}_\text{QMC}$ is its uncertainty, for $\rs^{(i)} = 1,2,3,4,5$.
$\widetilde{\chi}_\text{approx} = \chi_s^\text{approx}/\chi_s^{(0)}$ is computed using Eqs. (\ref{eq:chi_enh}) and (\ref{eq:alpha_c_pw92}).
Using the Perdew-Zunger \cite{perdew1981} ansatz for the spin-dependence of the correlation energy,
\begin{align}
  \varepsilon\suc(\rs,\zeta) &= \varepsilon\suc(\rs,0)
    + f(\zeta)\left[ \varepsilon\suc(\rs,1)
    - \varepsilon\suc(\rs,0) \right], \\
  f(\zeta) &= \frac{(1 + \zeta)^{4/3} + (1 - \zeta)^{4/3} - 2}{2^{4/3} - 2},
\end{align}
we have approximated
\begin{align}
  \alpha\suc^\text{QMC}(\rs) &= f''(0)\left[
    \varepsilon\suc^\text{QMC}(\rs,1)
  - \varepsilon\suc^\text{QMC}(\rs,0) \right], \\
  \delta \alpha\suc^\text{QMC}(\rs) &= f''(0)\left\{
    \left[\delta \varepsilon\suc^\text{QMC}(\rs,1)\right]^2
    + \left[\delta \varepsilon\suc^\text{QMC}(\rs,0)\right]^2
    \right\}^{1/2},
\end{align}
with $\varepsilon\suc^\text{QMC}$ the accurate correlation energies from Table VI of Ref. \cite{azadi2022}, and $\delta \varepsilon\suc^\text{QMC}$ their uncertainties.
A few values of the spin susceptibility enhancement predicted by QMC, PW92, and the present work are presented in Table \ref{tab:chi_enh}.

\clearpage
\twocolumngrid

\section{Evaluation of the fit quality at all values of $\rs$}

\begin{figure}[h]
  \centering
  \includegraphics[width=\columnwidth]{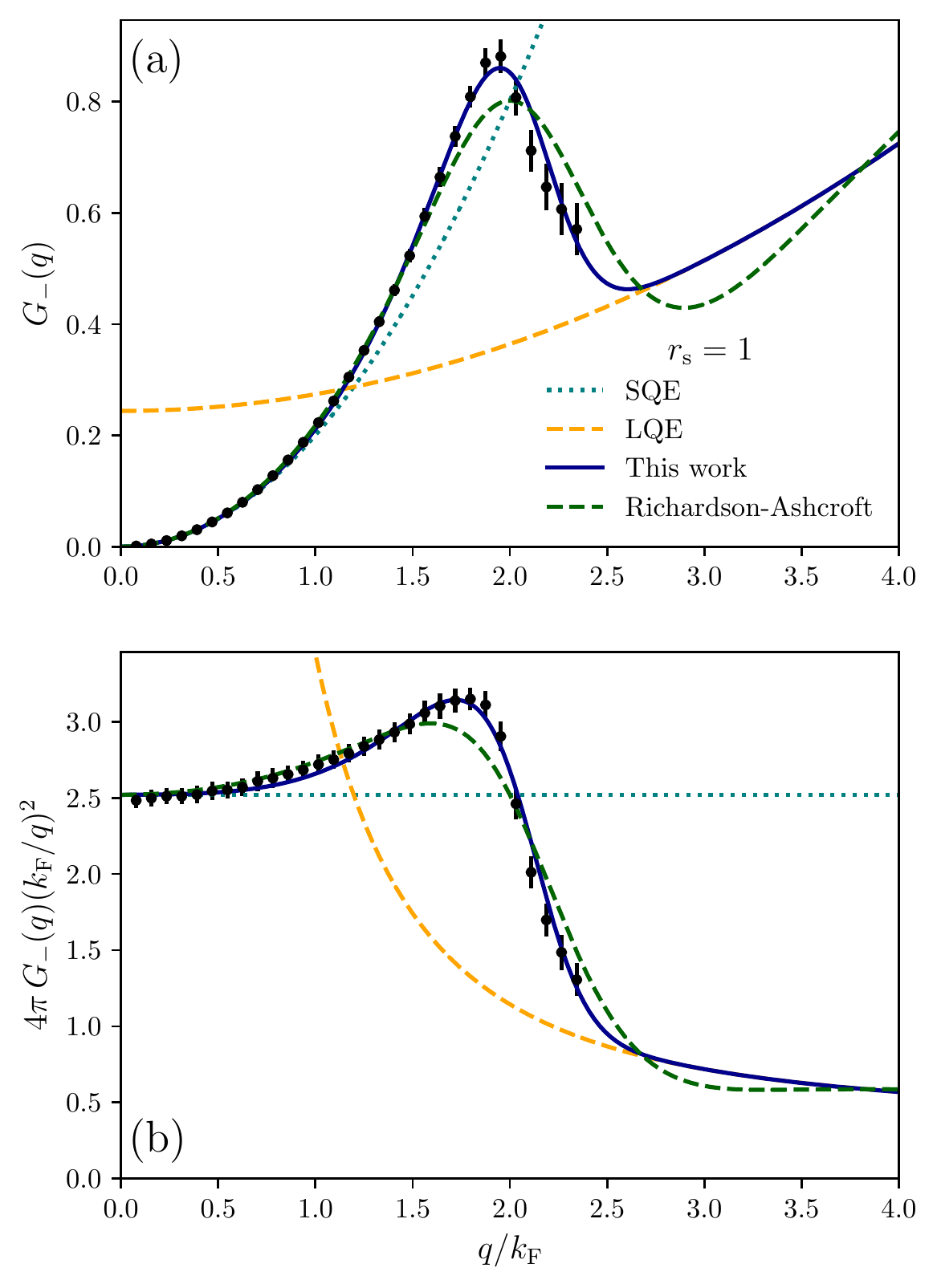}
  \caption{
  Comparison of the model $G_-$ of Eq. (\ref{eq:g_app}) (blue, solid curve) and Table \ref{tab:fpars} with the QMC data of Ref. \cite{kukkonen2021} (black points with vertical uncertainties) for $\rs = 1$.
  Panel (a) presents $G_-$ and (b) $4\pi G_- (\kf/q)^2 = \kf^2 f\suxc(q)$.
  The RA expression for $G_-$ \cite{richardson1994} (green, dashed), the small-$q$ expansion (SQE) of Eq. (\ref{eq:gp_small_q}) (teal, dotted), and large-$q$ expansion (LQE) of Eq. (\ref{eq:gp_large_q}) (orange, dashed) are also shown.
  }
  \label{fig:gm_rs_1}
\end{figure}

\begin{figure}
  \centering
  \includegraphics[width=\columnwidth]{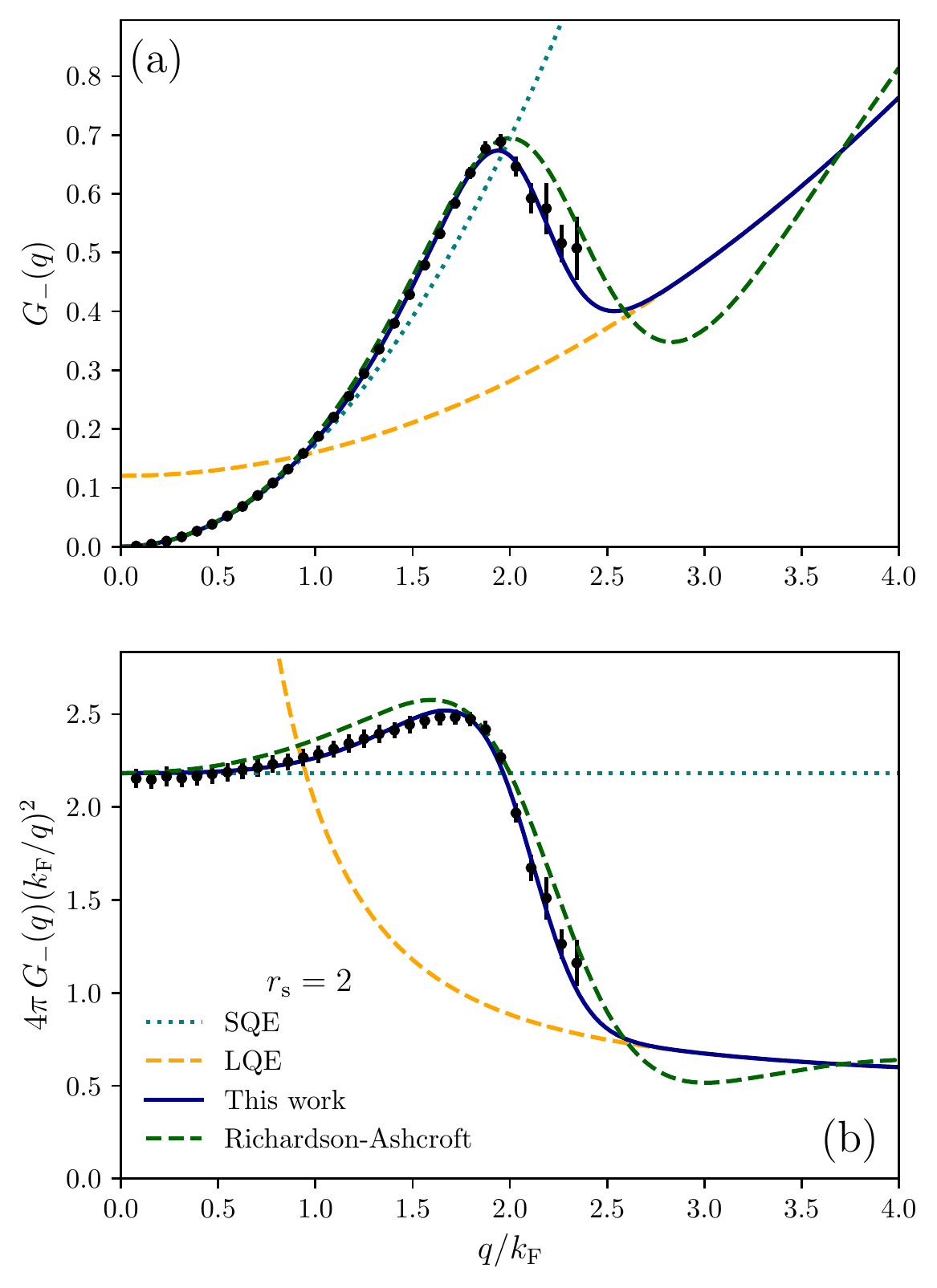}
  \caption{
  Same as Fig. \ref{fig:gm_rs_1}, but for $\rs = 2$.
  }
  \label{fig:gm_rs_2}
\end{figure}

\begin{figure}
  \centering
  \includegraphics[width=\columnwidth]{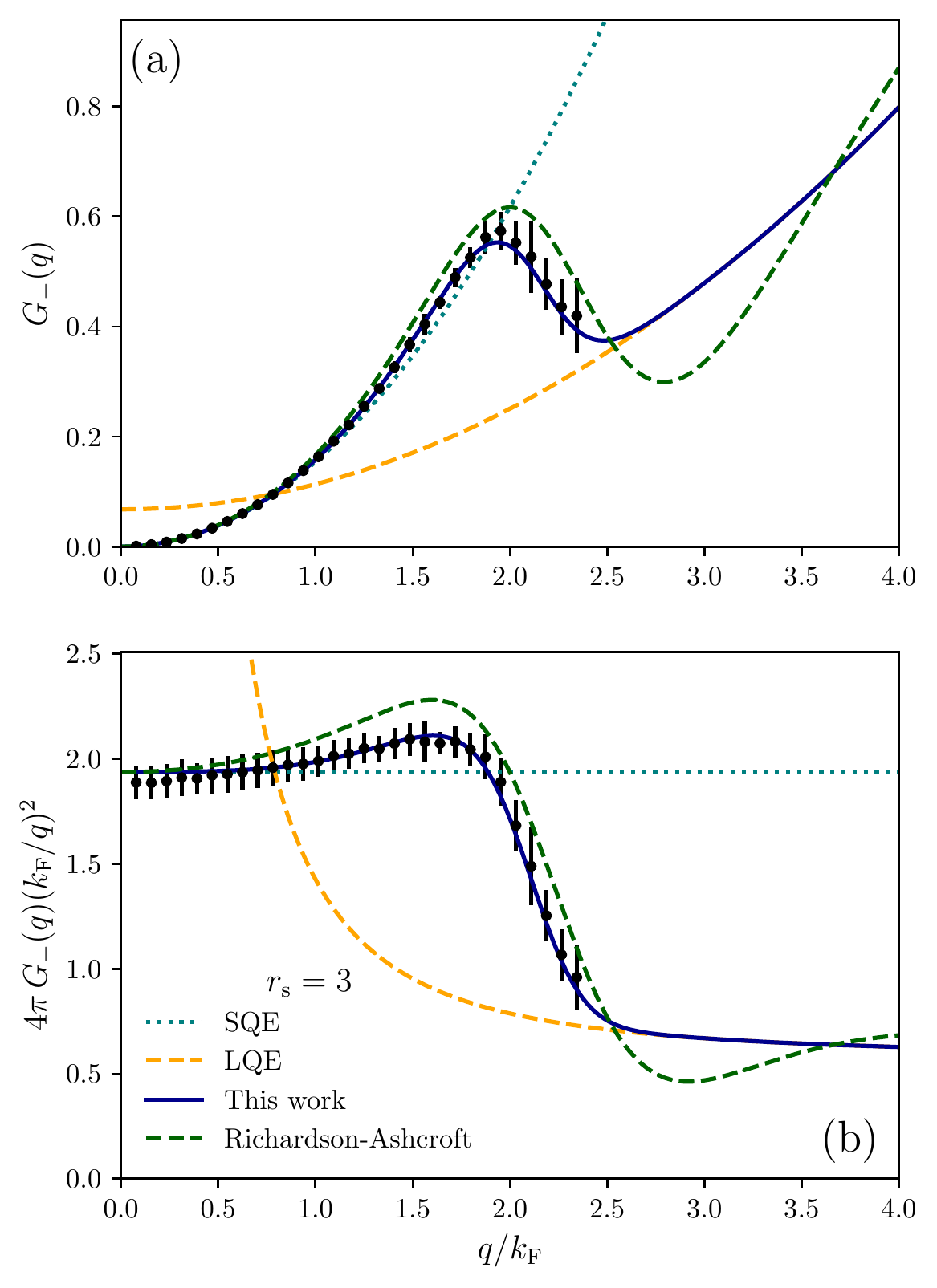}
  \caption{
  Same as Fig. \ref{fig:gm_rs_1}, but for $\rs = 3$.
  }
  \label{fig:gm_rs_3}
\end{figure}

\begin{figure}
  \centering
  \includegraphics[width=\columnwidth]{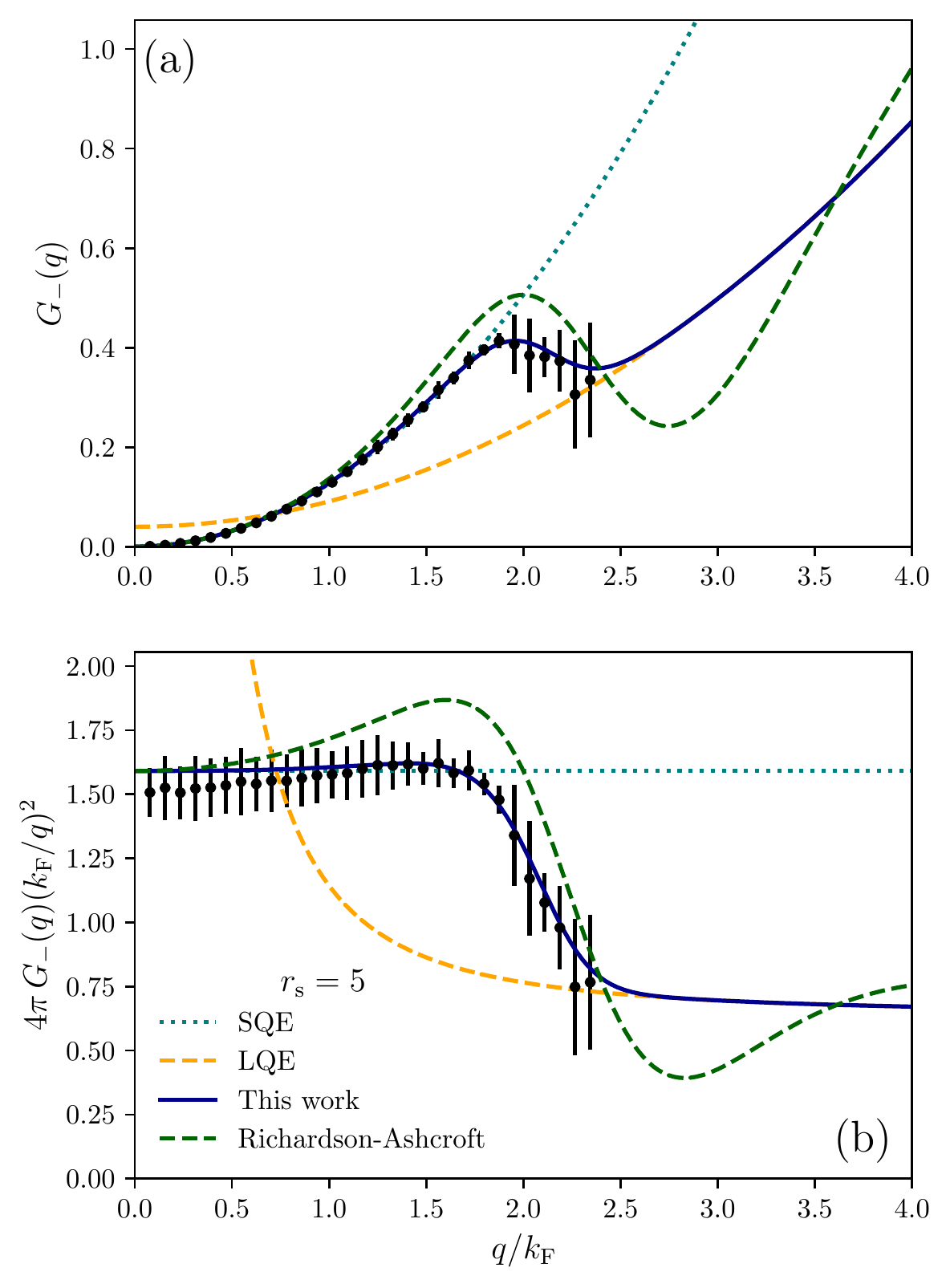}
  \caption{
  Same as Fig. \ref{fig:gm_rs_1}, but for $\rs = 5$.
  }
  \label{fig:gm_rs_5}
\end{figure}

This section presents figures analogous to Figs. \ref{fig:gplus_rs_2} and \ref{fig:gminus_rs_4} of the main text, but for the other values of $\rs$ used to fit $G_\pm(q)$.
For $G_+(\rs,q)$, these are for $\rs \in \{ 1, 5, 10 \}$ in Figs. \ref{fig:gp_rs_1}--\ref{fig:gp_rs_10}.
For $G_-(\rs,q)$, these are for $\rs \in \{ 1,2,3,5\}$ in Figs. \ref{fig:gm_rs_1}--\ref{fig:gm_rs_5}.

\subsection{Static density local field factor}

\nopagebreak
\begin{figure}[h]
  \centering
  \includegraphics[width=\columnwidth]{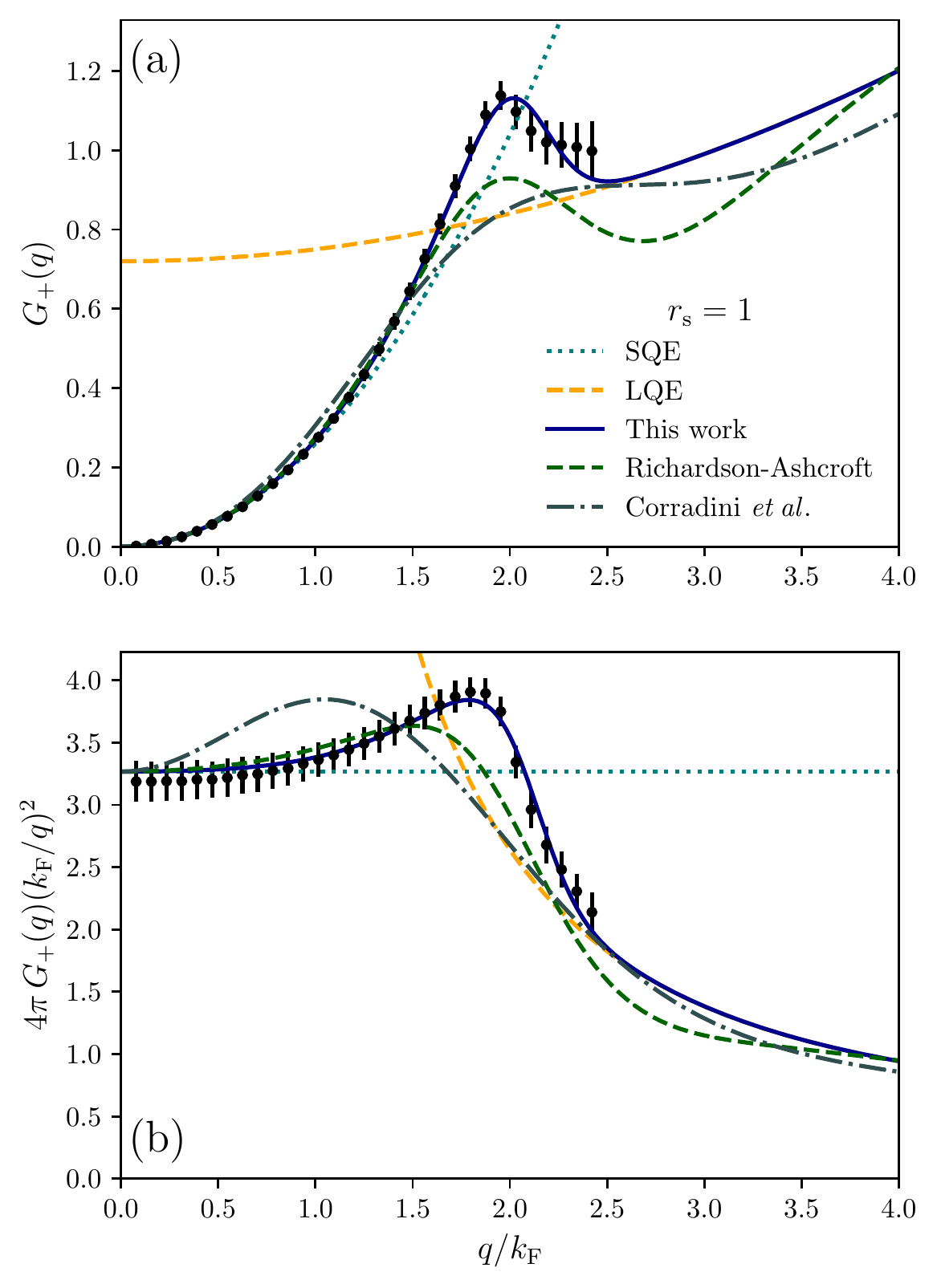}
  \caption{
  Comparison of the model $G_+$ of Eq. (\ref{eq:g_app}) (blue, solid curve) and Table \ref{tab:fpars} with the QMC data of Ref. \cite{kukkonen2021} (black circles with vertical uncertainties) for $\rs = 1$.
  Panel (a) presents $G_+$ and (b) $4\pi G_+ (\kf/q)^2 = \kf^2 f\suxc(q)$.
  Also shown are the LFFs of Corradini \textit{et al.} \cite{corradini1998} (gray, dash-dotted), which is fitted to the data of Ref. \cite{moroni1995}, and of RA \cite{richardson1994} (green, dashed).
  The small-$q$ expansion (SQE) of Eq. (\ref{eq:gp_small_q}) (teal, dotted) and large-$q$ expansion (LQE) of Eq. (\ref{eq:gp_large_q}) (orange, dashed) are also shown.
  }
  \label{fig:gp_rs_1}
\end{figure}

\begin{figure}[h]
  \centering
  \includegraphics[width=\columnwidth]{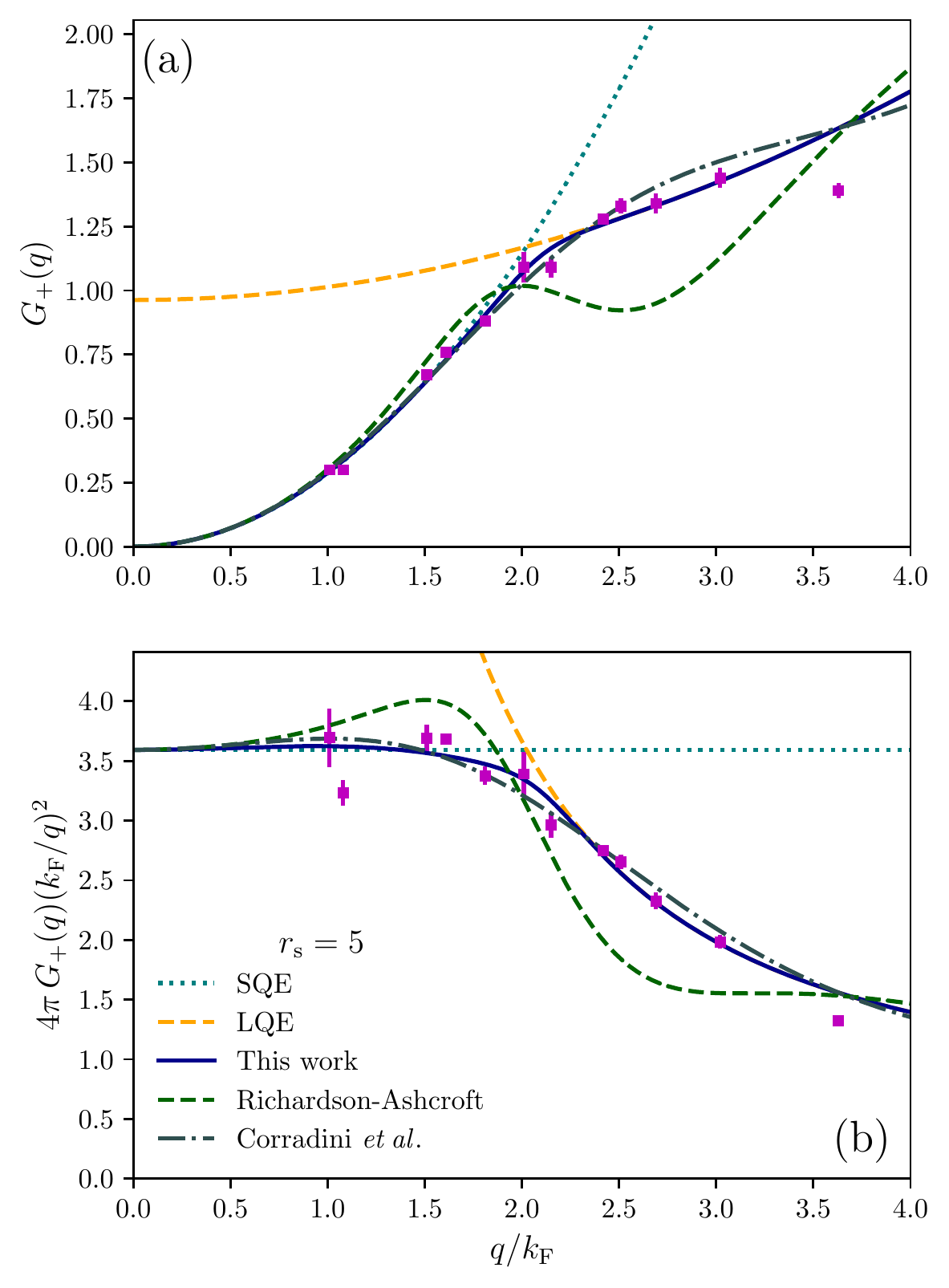}
  \caption{
  Comparison of the model $G_+$ of Eq. (\ref{eq:g_app}) (blue, solid curve) and Table \ref{tab:fpars} with the QMC data of Ref. \cite{moroni1995} (magenta squares with vertical uncertainties) for $\rs=5$.
  Panel (a) presents $G_+$ and (b) $4\pi G_+ (\kf/q)^2 = \kf^2 f\suxc(q)$.
  Also shown are the LFFs of Corradini \textit{et al.} \cite{corradini1998} (gray, dash-dotted), which is fitted to the data of Ref. \cite{moroni1995}, and of RA \cite{richardson1994} (green, dashed).
  The small-$q$ expansion (SQE) of Eq. (\ref{eq:gp_small_q}) (teal, dotted) and large-$q$ expansion (LQE) of Eq. (\ref{eq:gp_large_q}) (orange, dashed) are also shown.
  }
  \label{fig:gp_rs_5}
\end{figure}

\begin{figure}[h]
  \centering
  \includegraphics[width=\columnwidth]{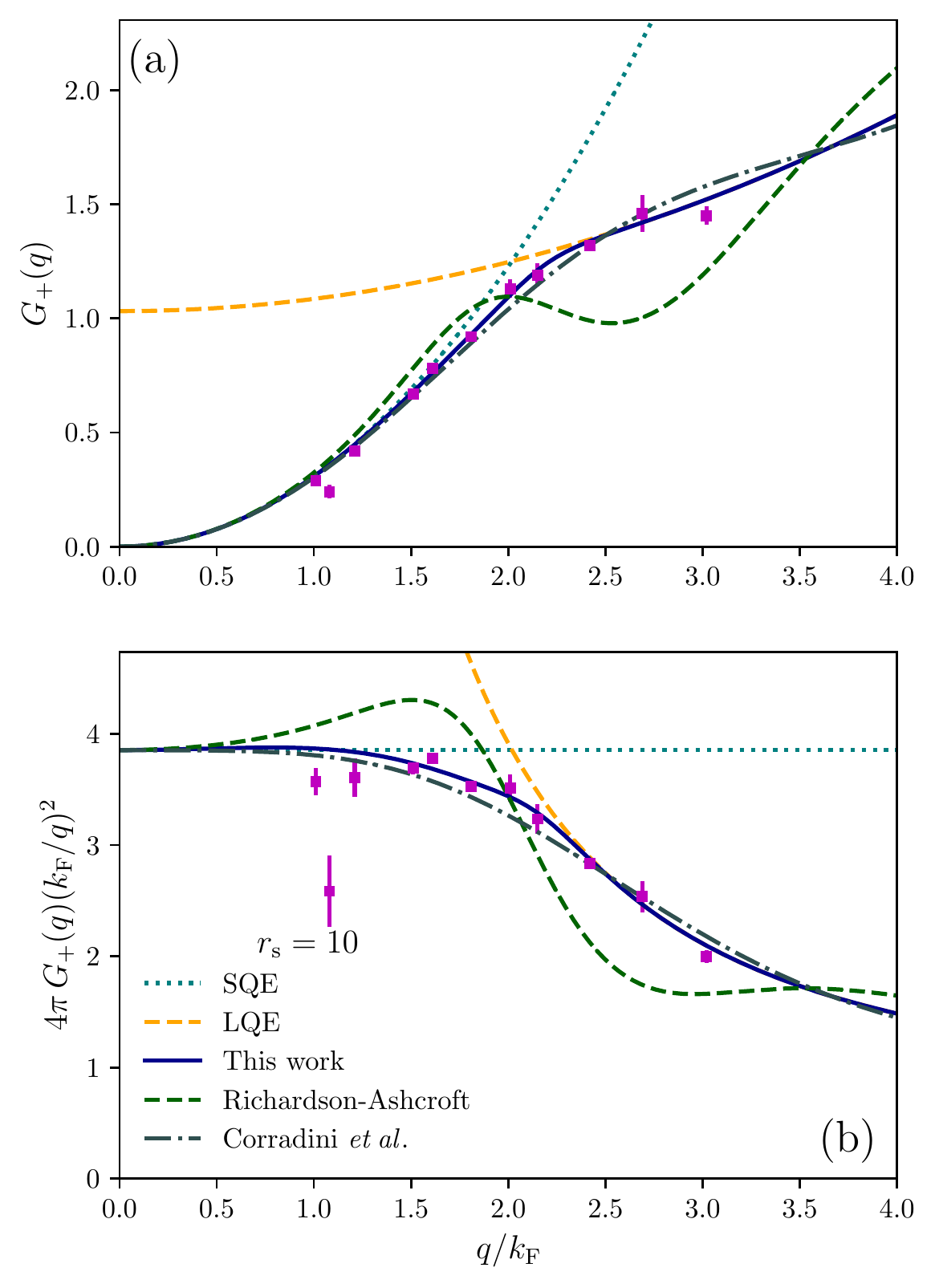}
  \caption{
  Same as Fig. \ref{fig:gp_rs_5}, but for $\rs = 10$.
  }
  \label{fig:gp_rs_10}
\end{figure}

\clearpage
\subsection{Static spin local field factor}

\nopagebreak

\clearpage

\section{Quality of extrapolation}

This section presents the \textit{predictions} of the model LFFs for the shapes of $G_\pm(q)$ at values of $\rs$ for which they are not fitted.
This gauges the quality of extrapolation and reliability of this model for jellium at any density.

\begin{figure}[h]
  \centering
  \includegraphics[width=\columnwidth]{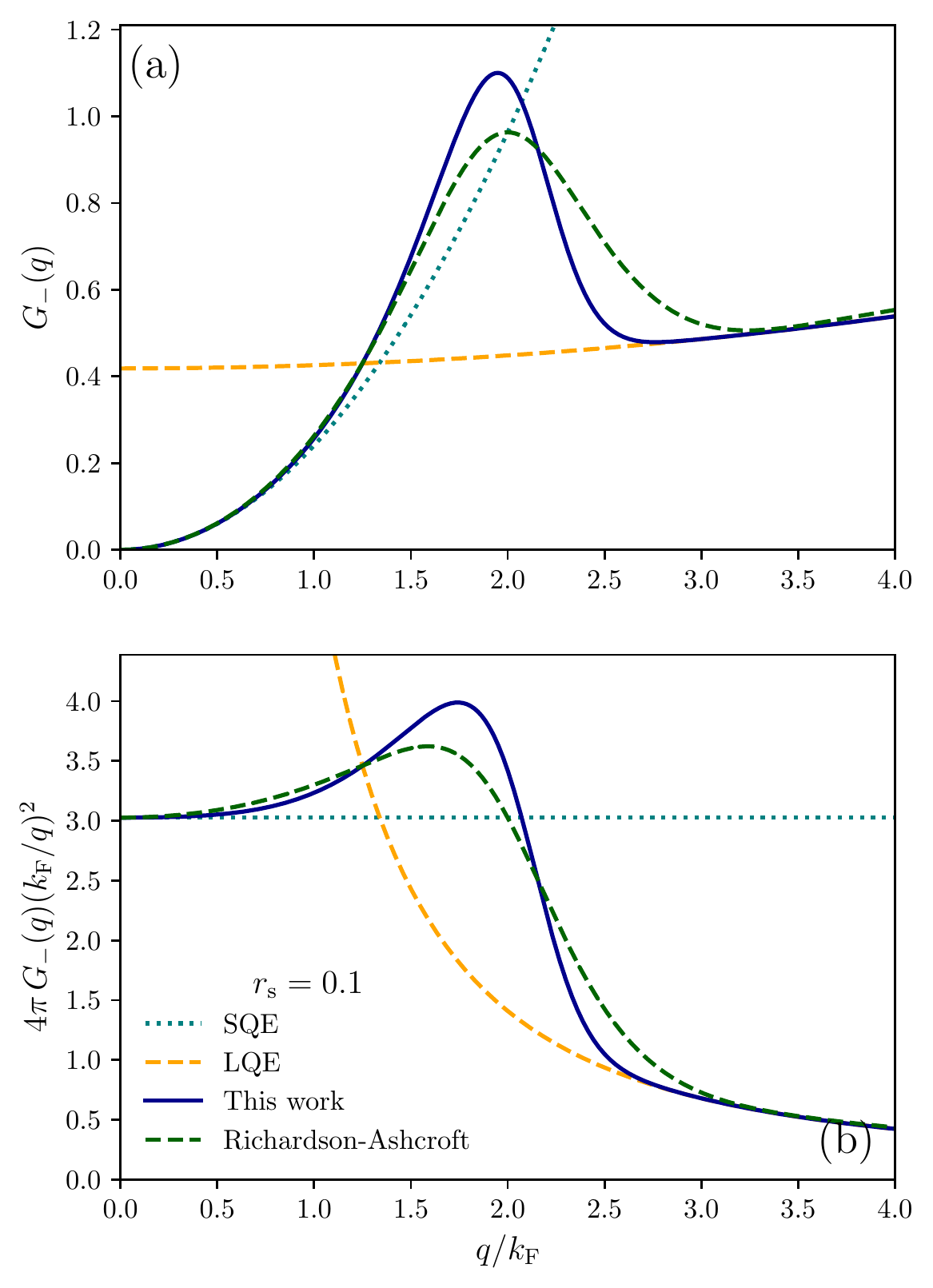}
  \caption{
  Extrapolation of the model $G_-$ to $\rs = 0.1$.
  Panel (a) presents $G_-$ and (b) $4\pi G_- (\kf/q)^2 = \kf^2 f\suxc(q)$.
  The RA expression for $G_-$ \cite{richardson1994} (red, dashed), the small-$q$ expansion (SQE) of Eq. (\ref{eq:gp_small_q}) (orange, dotted), and large-$q$ expansion (LQE) of Eq. (\ref{eq:gp_large_q}) (green, dotted) are also shown.
  }
  \label{fig:gm_rs_0p1}
\end{figure}

\begin{figure}[h]
  \centering
  \includegraphics[width=\columnwidth]{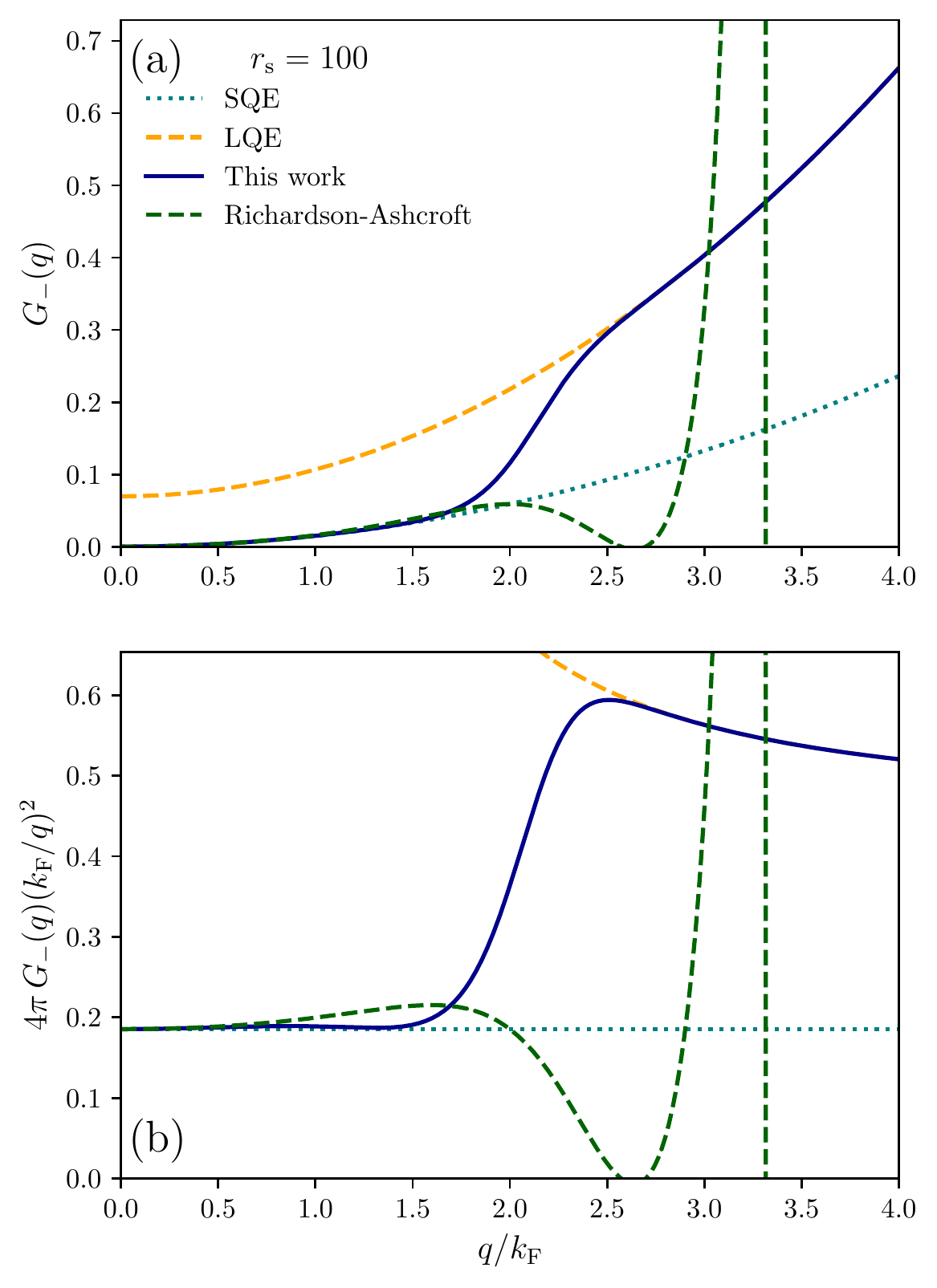}
  \caption{
  Same as Fig. \ref{fig:gm_rs_0p1}, but for $\rs = 100$.
  }
  \label{fig:gm_rs_100}
\end{figure}

For both $G_\pm(q)$, we show extrapolations to an extremely high density, $\rs = 0.1$ in Figs. \ref{fig:gp_rs_0p1} and \ref{fig:gm_rs_0p1}, and to an extremely low density, $\rs = 100$ in Figs. \ref{fig:gp_rs_100} and \ref{fig:gm_rs_100}.

\subsection{Static density local field factor}

\begin{figure}[h]
  \centering
  \includegraphics[width=\columnwidth]{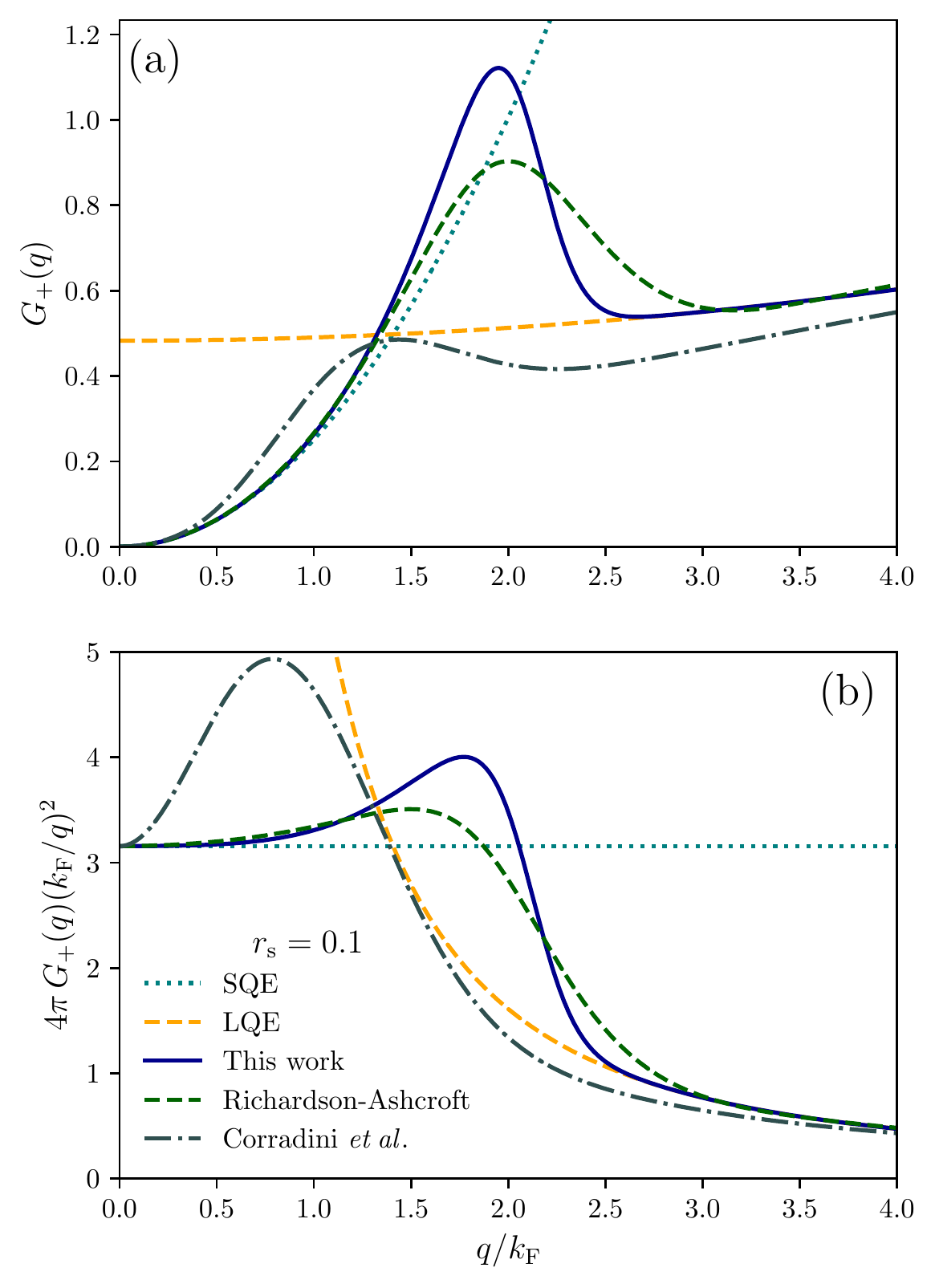}
  \caption{
  Extrapolation of the model $G_+$ to $\rs=0.1$.
  Panel (a) presents $G_+$ and (b) $4\pi G_+ (\kf/q)^2 = \kf^2 f\suxc(q)$.
  Also shown are the LFFs of Corradini \textit{et al.} \cite{corradini1998} (gray, dash-dotted), which is fitted to the data of Ref. \cite{moroni1995}, and of RA \cite{richardson1994} (red, dashed).
  The small-$q$ expansion (SQE) of Eq. (\ref{eq:gp_small_q}) (orange, dotted) and large-$q$ expansion (LQE) of Eq. (\ref{eq:gp_large_q}) (green, dotted) are also shown.
  }
  \label{fig:gp_rs_0p1}
\end{figure}

\begin{figure}[h]
  \centering
  \includegraphics[width=\columnwidth]{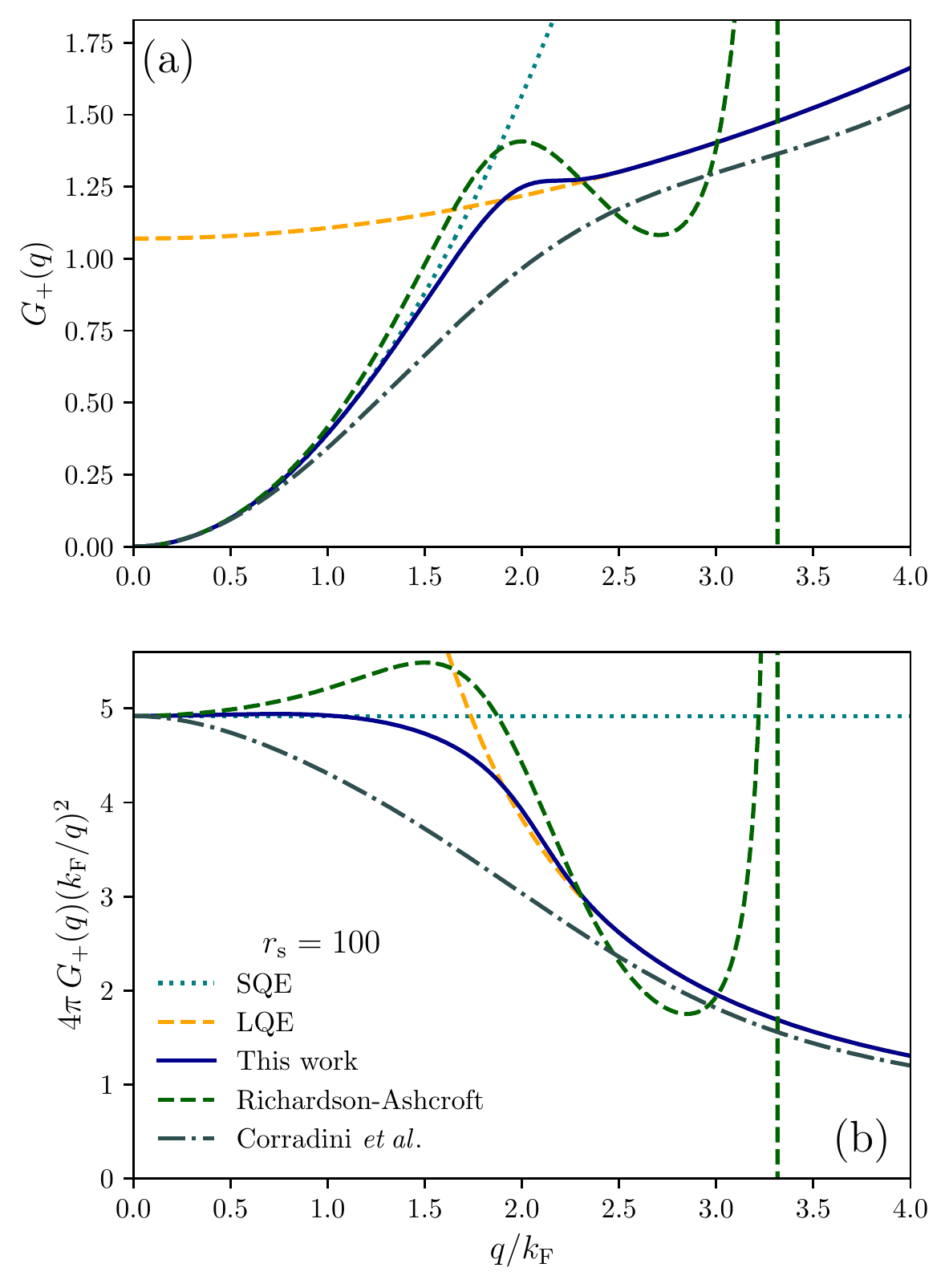}
  \caption{
  Same as Fig. \ref{fig:gp_rs_0p1}, but for $\rs = 100$.
  }
  \label{fig:gp_rs_100}
\end{figure}

\clearpage
\subsection{Static spin local field factor}

\clearpage
\onecolumngrid

\section{Surface plots of the local field factors}

\begin{figure}[h]
  \includegraphics[width=0.8\columnwidth]{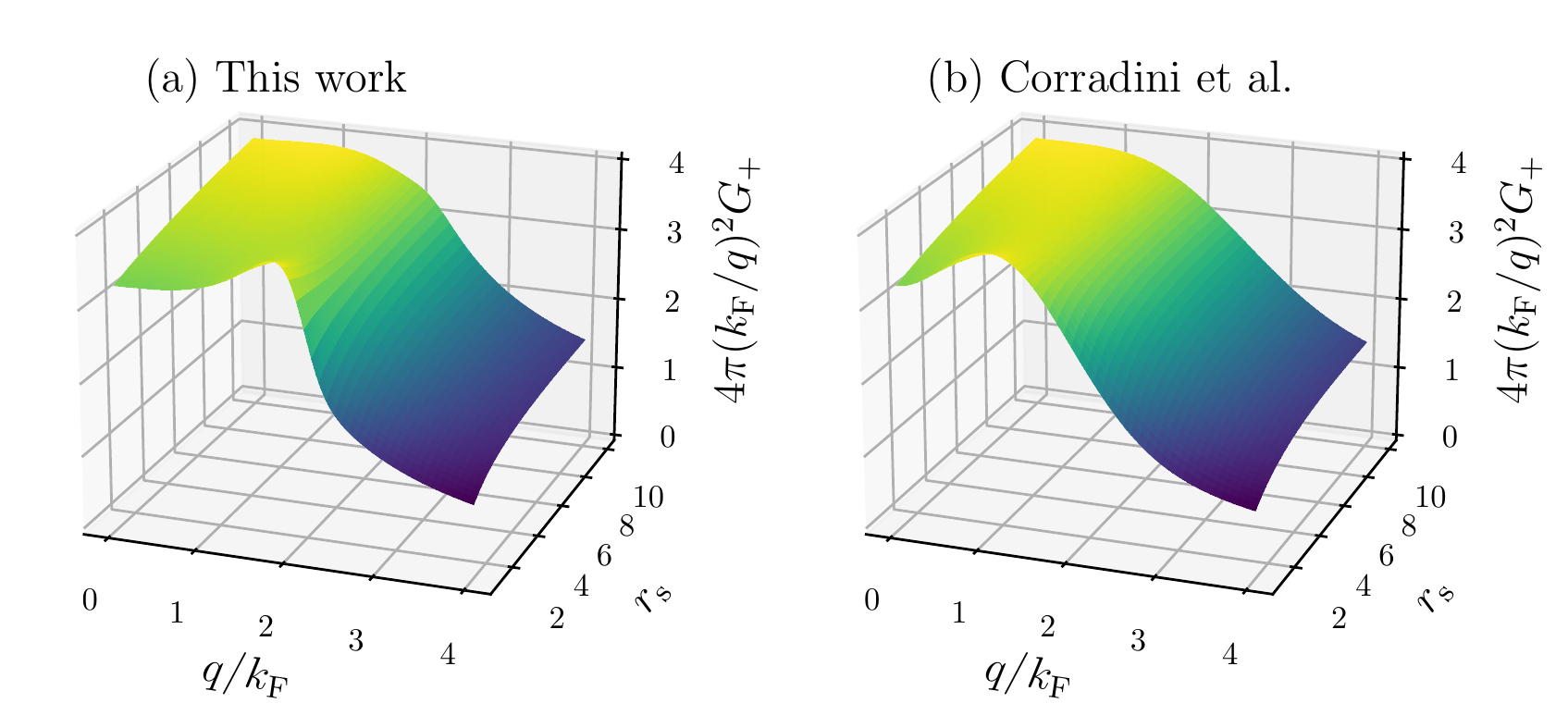}
  \caption{Surface plot of (a) the model $4\pi G_+(\rs,q)(\kf/q)^2$ of this work and (b) of Corradini \textit{et al.} \cite{corradini1998}.
  Both are shown as functions of $0 \leq q/\kf \leq 4$ and in the metallic range $2 \leq \rs \leq 10$.
  }
  \label{fig:surf_gp_corr}
\end{figure}

\begin{figure}[h]
  \includegraphics[width=0.8\columnwidth]{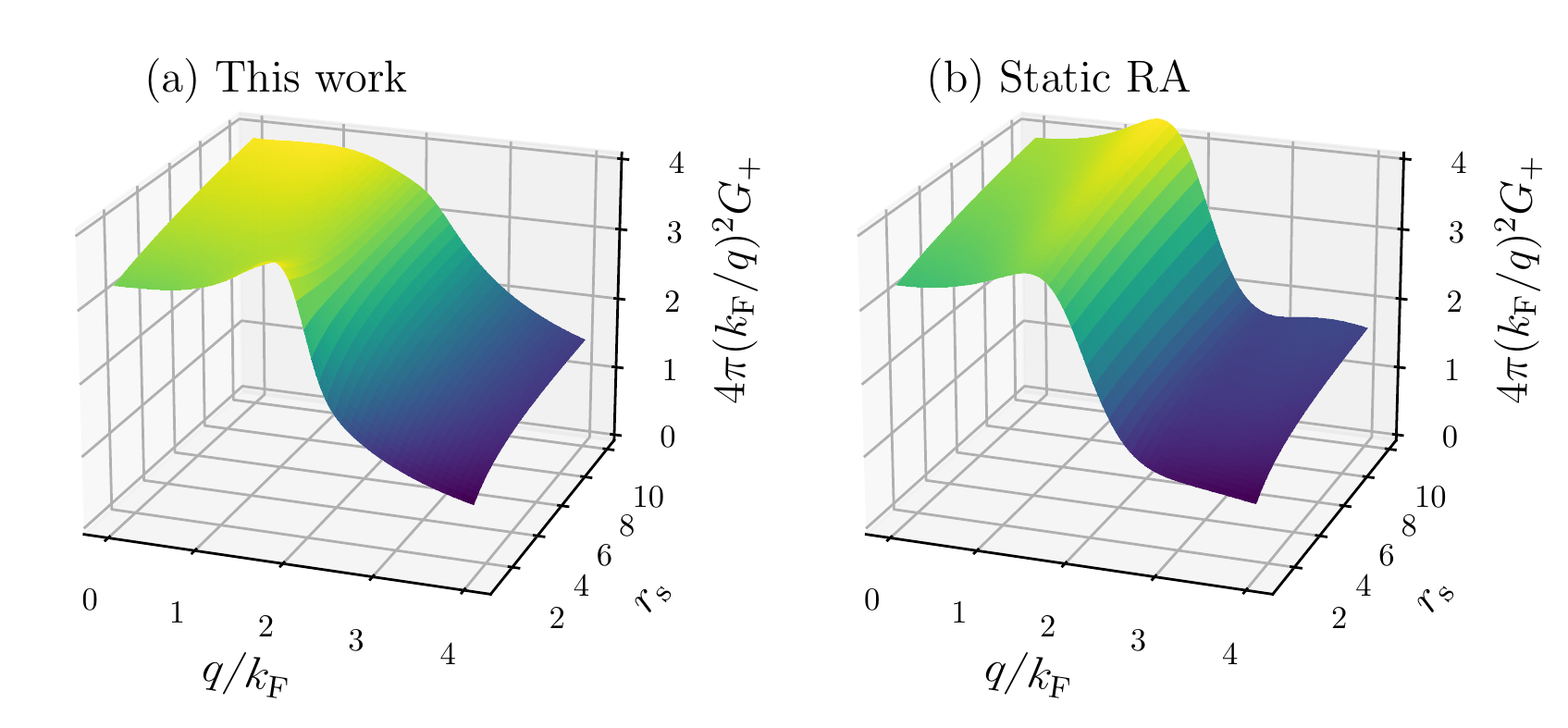}
  \caption{Surface plot of (a) the model $4\pi G_+(\rs,q)(\kf/q)^2$ of this work and (b) of Richardson and Ashcroft (RA) \cite{richardson1994}.
  Both are shown as functions of $0 \leq q/\kf \leq 4$ and in the metallic range $2 \leq \rs \leq 10$.
  }
  \label{fig:surf_gp_ras}
\end{figure}

\begin{figure}[h]
  \includegraphics[width=0.8\columnwidth]{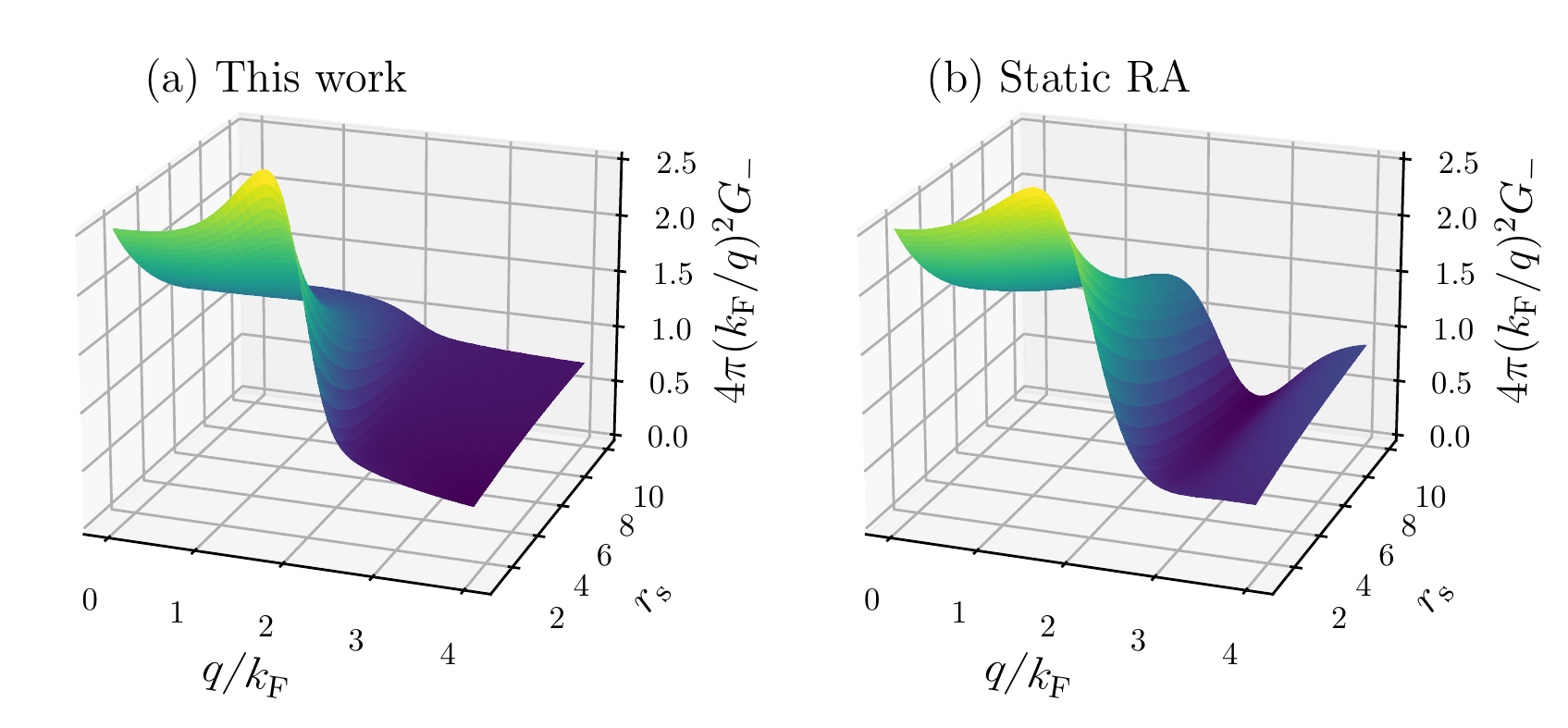}
  \caption{Surface plot of (a) the model $4\pi G_-(\rs,q)(\kf/q)^2$ of this work and (b) of Richardson and Ashcroft (RA) \cite{richardson1994}.
  Both are shown as functions of $0 \leq q/\kf \leq 4$ and in the metallic range $2 \leq \rs \leq 10$.
  }
  \label{fig:surf_gm_ras}
\end{figure}

This section presents surface plots of $G_+(\rs,q)$ as a function of $q/\kf$ and $\rs$, with comparisons to the Corradini \textit{et al.} LFF in Fig. \ref{fig:surf_gp_corr}, and to the Richardson-Ashcroft (RA) LFF in Fig. \ref{fig:surf_gp_ras}.
The model of $G_-(\rs,q)$ developed here and the model of RA are compared in Fig. \ref{fig:surf_gm_ras}.

\section{Computation of correlation energies \label{sec:eps_c_method}}

To compute correlation energies per electron for a spin-unpolarized jellium, $\varepsilon\suc(\rs,\zeta=0)$, we use the standard coupling-constant integration \cite{lein2000}
\begin{equation}
  \varepsilon\suc(\rs,\zeta=0) = -3 \int_0^\infty d \left(\frac{q}{\kf} \right) \int_0^1 d \lambda \int_0^\infty
    d \left( \frac{u}{\kf^2} \right) \frac{\left[\chi_0(q,i u) \right]^2 f_\mathrm{Hxc}^{(\lambda)}(q,i u)}{1 - \chi_0(q,i u) f_\mathrm{Hxc}^{(\lambda)}(q,i u)}.
    \label{eq:acfdt}
\end{equation}
$\chi_0$ is the non-interacting or Kohn-Sham response function.
When evaluated for the UEG, it also known as the Lindhard function \cite{lindhard1954},
\begin{equation}
  \chi_0(q,i u) = \frac{\kf}{2\pi^2} \left\{
    \frac{z^2 - U^2 - 1}{4 z} \ln \left[\frac{U^2 + (z + 1)^2}{U^2 + (z - 1)^2} \right]
    -1 + U \arctan \left(\frac{1 + z}{U} \right)
    + U \arctan \left(\frac{1 - z}{U} \right)
  \right\},
\end{equation}
where $z = q/(2 \kf)$ and $U \equiv u/(q\kf)$.
$f_\mathrm{Hxc}^{(\lambda)}$ is the sum of Hartree,
\begin{equation}
  f_\mathrm{H}(q) = \frac{4\pi}{q^2},
\end{equation}
and exchange-correlation kernels evaluated at the coupling-constant $\lambda$.
From Ref. \cite{lein2000}, we may obtain this expression from the coupling-constant scaled LFF
\begin{equation}
  f_\mathrm{Hxc}^{(\lambda)}(q,i u) = \frac{4\pi \lambda}{q^2} \left[1 - G_+\left(\lambda \rs, \frac{q}{\lambda},\frac{i u}{\lambda^2} \right) \right].
  \label{eq:fhxc}
\end{equation}

Developing a method to reliably perform the three-dimensional integration needed in Eq. (\ref{eq:acfdt}) without combinatorial explosion is challenging.
To do this, we first computed approximate random phase approximation (RPA) correlation energies by integrating up to two cutoffs, called $x\suc \equiv q\suc/\kf$ and $v\suc \equiv u\suc/\kf^2$,
\begin{equation}
  \varepsilon\suc^\text{RPA}(\rs) \approx -3 \int_0^{x\suc} dx \int_0^1 d \lambda \int_0^{v\suc}
    d v \frac{4\pi \lambda\left[\chi_0(q,i u) \right]^2 q^{-2}}{1 - 4\pi \lambda \chi_0(q,i u)/q^{-2}}.
  \label{eq:eps_c_rpa_approx}
\end{equation}
As $G_+^\text{RPA}=0$, the right- and left-hand-sides of Eq. (\ref{eq:eps_c_rpa_approx}) become exactly equal in the limit that $x\suc, \, v\suc \to \infty$.
These integrals were computed using globally-adaptive, Gauss-Kronrod quadrature.
See the computational details of Refs. \cite{perdew2021} and \cite{kaplan2022} for more details.

The cutoffs were adjusted to give agreement to within, ideally, 1\% error of the PW92-parameterized RPA correlation energies \cite{perdew1992}.
These cutoffs were then approximately parameterized as continuous functions of $\rs$,
\begin{equation}
  x\suc(\rs) \approx \left\{
  \begin{array}{ll}
    c_{x0} + c_{x1} \rs, & \rs \leq 5 \\
    c_{x0} + 5 c_{x1} + c_{x2} (\rs - 5) + c_{x3} (\rs - 5)^2, & 5 < \rs \leq 60 \\
    c_{x0} + 5 c_{x1} + 55 c_{x2} + 3025 c_{x3} + c_{x4}(\rs - 60), & 60 < \rs
  \end{array}
  \right. ,
\end{equation}
with $c_{x0} = 3.928319$, $c_{x1} = 0.540168$, $c_{x2} = 0.042225$, $c_{x3} = 0.001810$, and $c_{x4} =2.501585$.
Analogously,
\begin{equation}
  v\suc(\rs) \approx \left\{
  \begin{array}{ll}
    c_{v0} + c_{v1} \rs^{c_{v2}}, & \rs \leq 40 \\
    c_{v0} + c_{v1}(40)^{c_{v2}} + (\rs - 40)^{c_{v3}}, & 40 < \rs
  \end{array}
  \right. ,
\end{equation}
with $c_{v0} = 1.227277$, $c_{v1} = 5.991171$, $c_{v2} = 0.283892$, and $c_{v3} = 0.379981$.

To recover the error lost in using finite integration bounds, we then perform a set of coordinate remappings.
Let $f(x)$ be a generic function of $x$, and $g(v)$ a generic function of $v$.
Then the mappings used are
\begin{align}
  \int_0^\infty dx \, f(x) &= \int_0^{x\suc} dx \, f(x) + \int_0^{1/x\suc} dt \frac{f(1/t)}{t^2} \\
  \int_0^\infty dv \, g(v) &= \int_0^{v\suc} dv \, g(v) + \int_0^1 dw \frac{g(v\suc - \ln(1 - w))}{1 - w}.
\end{align}
These mappings are, in principle, exact.
For the range of $0 < x < x\suc$, we use 100-point Gauss-Legendre quadrature, and for the range of $0 < t < 1/x\suc$, we use 50-point Gauss-Legendre quadrature.
The same number of points were used for the corresponding ranges of $v$ and $w$, respectively.
100-point Gauss-Legendre quadrature was used for the coupling-constant, $\lambda$, integration.
Table \ref{tab:RPA_sanity} shows that this method becomes asymptotically exact as $\rs \to 0$, and, in the metallic range $ 1 \leq \rs \leq 10$, gives generally negligible percent deviations from the Perdew-Wang parameterization of the RPA correlation energy, PW-RPA \cite{perdew1992}.
Indeed, for all $\rs \leq 120$, this method yields percent deviations less than 1\% from PW-RPA.

\begin{table}
  \centering
  \begin{tabular}{rrrr} \hline
    $\rs$ & $\varepsilon\suc^\mathrm{RPA}(\rs)$ & $\varepsilon\suc^\mathrm{PW-RPA}(\rs)$ & Percent Deviation (\%) \\ \hline
    0.1 & -0.143815 & -0.143819 & 0.00 \\
    0.5 & -0.097155 & -0.097221 & 0.07 \\
    1.0 & -0.078631 & -0.078741 & 0.14 \\
    2.0 & -0.061651 & -0.061797 & 0.24 \\
    3.0 & -0.052619 & -0.052774 & 0.29 \\
    4.0 & -0.046673 & -0.046827 & 0.33 \\
    5.0 & -0.042343 & -0.042491 & 0.35 \\
    10.0 & -0.030549 & -0.030661 & 0.37 \\
    20.0 & -0.021288 & -0.021367 & 0.37 \\
    40.0 & -0.014385 & -0.014454 & 0.48 \\
    60.0 & -0.011300 & -0.011367 & 0.59 \\
    80.0 & -0.009472 & -0.009542 & 0.74 \\
    100.0 & -0.008236 & -0.008311 & 0.90 \\
    120.0 & -0.007345 & -0.007413 & 0.93 \\
  \hline
  \end{tabular}
  \caption{Comparison of the RPA correlation energies computed using the method described here, and with the Perdew-Wang approximation for the RPA correlation energy, PW-RPA \cite{perdew1992}.
  The PW-RPA approximation is simply a parameterization of the accurate RPA data of Vosko, Wilk, and Nusair \cite{vosko1980}.
  Percent deviations, $100\% \cdot (1 - \varepsilon\suc^\mathrm{RPA}/\varepsilon\suc^\mathrm{PW-RPA})$, are shown in the last column.
  }
  \label{tab:RPA_sanity}
\end{table}

\section{Corrected expressions for the Richardson-Ashcroft local field factors \label{sec:RA_corrected}}

The work of Richardson and Ashcroft \cite{richardson1994} is extremely important, as it is the first work to directly compute the individual LFFs $G_s$, $G_a$, and $G_n$ at a range of wavevectors, frequencies, and densities.
Moreover, they provided sensible parameterizations of these functions that are unfortunately hindered by typographical errors, as realized by Lein \textit{et al.} \cite{lein2000}.
We provide further corrections here.
The density and spin LFFs are computed as
\begin{align}
  G_+(\rs,q,\omega) &= G_s(\rs,q,\omega) + G_n(\rs,q,\omega) \\
  G_-(\rs,q,\omega) &= G_a(\rs,q,\omega) + G_n(\rs,q,\omega).
\end{align}
As before, $q>0$ is a wavevector, and $\omega$ is a complex-valued frequency.
The following dimensionless variables are used in the Richardson-Ashcroft work
\begin{align}
  z &= q/(2\kf) \\
  u &= \frac{1}{2\kf^2} \mathrm{Im} \, \omega.
\end{align}
A few $\rs$-dependent functions are used to define the low- and high-frequency regimes of the LFFs, $\lambda_i^{(j)}$, where $i= s, \, a, \, n$ and $j = 0, \, \infty$.
Richardson and Ashcroft parameterized the relationship between the $u \to 0$ behaviors of $G_a$ and $G_n$ as
\begin{equation}
  \frac{\lambda_n^{(0)}}{\lambda_n^{(0)} + \lambda_a^{(0)}}
    \approx \frac{-(0.11) \rs}{1 + (0.33) \rs} \equiv \mathcal{F}(\rs).
\end{equation}
Their sum is rigorously computed using Eq. (RA:39) of Ref. \cite{lein2000},
\begin{equation}
  \lambda_n^{(0)} + \lambda_a^{(0)} = 1 - 3 \left(\frac{2\pi}{3} \right)^{2/3}
    \rs \frac{\partial^2 \varepsilon\suc}{\partial \zeta^2} (\rs,0),
\end{equation}
where $\varepsilon\suc(\rs,\zeta)$ is in \textit{Hartree} units, and not Rydberg units as in Ref. \cite{richardson1994} or Eq. (RA:39) of Ref. \cite{lein2000}.
Thus
\begin{align}
  \lambda_n^{(0)} &= \mathcal{F}(\rs) \left[
    1 - 3 \left(\frac{2\pi}{3} \right)^{2/3}
      \rs \frac{\partial^2 \varepsilon\suc}{\partial \zeta^2} (\rs,0)
  \right], \\
  \lambda_a^{(0)} &= \frac{1 - \mathcal{F}(\rs)}{\mathcal{F}(\rs)} \lambda_n^{(0)}.
\end{align}
The $u \to 0$ limit of the spin-symmetric, noninteracting LFF is then
\begin{equation}
  \lambda_s^{(0)} = - \lambda_n^{(0)} + 1
  + \frac{2\pi}{3}a\sux \rs^2 \frac{\partial \varepsilon\suc}{\partial \rs}(\rs,0) - \frac{\pi}{3} a\sux \rs^3 \frac{\partial^2 \varepsilon\suc}{\partial \rs^2}(\rs,0)
\end{equation}
again with $\varepsilon\suc(\rs,\zeta)$ in Hartree.
$a\sux = [4/(9\pi)]^{1/3}$ is the inverse of the factor that relates the Fermi momentum to the Wigner-Seitz radius, $\rs = (a\sux \kf)^{-1}$.

Although not defined explicitly in Ref. \cite{richardson1994}, the high-frequency limit of the spin-antisymmetric, noninteracting LFF is
\begin{equation}
  \lambda_a^{(\infty)} = \frac{2g(\rs) - 1}{3},
\end{equation}
where again, $g(\rs)$ is the on-top pair distribution function.
The high-frequency limit of the occupation number LFF is given as
\begin{equation}
  \lambda_n^{(\infty)} = 6\pi a\sux \rs \frac{\partial }{\partial \rs}\left[
    \rs \, \varepsilon\suc(\rs,0) \right],
\end{equation}
and the corresponding limit of the spin-symmetric, noninteracting LFF from Eq. (RA:39) of Ref. \cite{lein2000},
\begin{equation}
  \lambda_s^{(\infty)} = \frac{3}{5} - \frac{4\pi a\sux}{5}\left[
    \rs^2 \frac{\partial \varepsilon\suc}{\partial \rs} (\rs,0)
    + 2\rs \varepsilon\suc(\rs,0) \right].
\end{equation}

Finally, we give the expression for the spin-symmetric, noninteracting LFF as
\begin{align}
  \gamma_s & \equiv \frac{9}{16[1 - g(\rs)]} \lambda_s^{(\infty)} + \frac{4\alpha_s - 3}{4\alpha_s} \\
  a_s(u) &= \lambda_s^{(\infty)}
    + \frac{\lambda_s^{(0)} - \lambda_s^{(\infty)}}{1 + (\gamma_s u)^2} \\
  c_s(u) &= \frac{3 \lambda_s^{(\infty)}}{4[ 1 - g(\rs) ]}
    - \left(
      1 + \gamma_s u \right)^{-1}
      \left[\frac{4}{3} - \frac{1}{\alpha_s}
      + \frac{3\lambda_s^{(\infty)}}{4[1 - g(\rs)]}
    \right] \\
  b_s(u) &= a_s(u) \left\{
    3 a_s(u)(1 + u)^4 - \frac{8}{3}[1 - g(\rs)](1 + u)^3 - 2c_s(u)[1 - g(\rs)](1 + u)^4
  \right\}^{-1} \\
  G_s(z,i u) &= z^2 \frac{a_s(u) + 2 [1 - g(\rs)] b_s(u)z^6/3}{1 + c_s(u)z^2 + b_s(u)z^8}.
\end{align}
$\alpha_s = 0.9$ is a fit parameter.

Likewise, the spin-antisymmetric, noninteracting LFF is parameterized as
\begin{align}
  \gamma_a &= \frac{9}{8} \lambda_a^{(\infty)} + \frac{1}{4}, \\
  a_a(u) &= \lambda_a^{(\infty)}
    + \frac{\lambda_a^{(0)} - \lambda_a^{(\infty)}}{1 + (\gamma_a u)^2} \\
  c_a(u) &= \frac{3}{2} \lambda_a^{(\infty)} - \left[ 1 + (\gamma_a u)^2 \right]^{-1}
    \left[ \frac{1}{3} + \frac{3}{2} \lambda_a^{(\infty)} \right] \\
  \beta_a(u) &= \frac{4g(\rs) - 1}{3}
    - \lambda_a^{(\infty)}\, \frac{(\gamma_a u)^2}{1 + (\gamma_a u)^2} \\
  b_a(u) &= a_a(u) \left[  3a_a(u)(1 + u)^4 - 4\beta_a(u)(1 + u)^3 - 3 c_a(u) \beta_a(u)(1 + u)^4
  \right]^{-1} \\
  G_a(z,i u) &= \lambda_a^{(\infty)} \frac{(\gamma_a u)^2}{1  + (\gamma_a u)^2}
     + z^2 \frac{a_a(u) + b_a(u) \beta_a(u) z^6}{1 + c_a(u) z^2 + b_a(u) z^8}.
\end{align}

Last, the occupation number LFF is parameterized as
\begin{align}
  a_n(u) &= \lambda_n^{(\infty)} + \frac{\lambda_n^{(0)} - \lambda_n^{(\infty)}}{1 + (\gamma_n u)^2} \\
  c_n(u) &= \frac{3 \gamma_n u}{(1.18)(1 + \gamma_n u)} - [1 + (\gamma_n u)^2]^{-1}
    \left[ \frac{ 3\lambda_n^{(0)} + \lambda_n^{(\infty)} }{3 \lambda_n^{(0)} + 2\lambda_n^{(\infty)} } + \frac{3\gamma_n u}{(1.18)(1 + \gamma_n u)} \right] \\
  d_n(u) &= a_n(u) + \lambda_n^{(\infty)} + \frac{2}{3} \lambda_n^{(\infty)} c_n(u)(1 + \gamma_n u) \\
  b_n(u) &= -\frac{3}{2\lambda_n^{(\infty)}(1 + \gamma_n u)^2}\left\{
    d_n(u) + \left[ d_n(u)^2 + \frac{4}{3} \lambda_n^{(\infty) } a_n(u) \right]^{1/2}
  \right\} \\
  G_n(z, i u) &= z^2 \frac{a_n(u) - \lambda_n^{(\infty)} b_n(u) z^4/3}{1 + c_n(u) z^2 + b_n(u) z^4}.
\end{align}
$\gamma_n = 0.68$ is another fit parameter.

\end{document}